\documentclass[10pt,twocolumn,twoside]{IEEEtran}
\usepackage{pifont}
\usepackage{bbding}  

\IEEEoverridecommandlockouts                              
\usepackage{amsmath,amssymb,epsf,epsfig,times}
\usepackage{amsfonts}
\usepackage[all]{xy}
\usepackage{color}
\usepackage{subfigure}
\usepackage{url,cite}

\newtheorem{theorem}{Theorem}[section]
\newtheorem{lemma}{Lemma}[section]
\def\proof{\noindent{\it Proof: }}
\def\QED{\mbox{\rule[0pt]{1.5ex}{1.5ex}}}
\def\endproof{\hspace*{\fill}~\QED\par\endtrivlist\unskip}
\newcommand{\re}{\mathbb{R}}

\newcommand{\defeq}{\stackrel{\triangle}{=}}

\newtheorem{definition}[theorem]{Definition}

\newtheorem{remark}[theorem]{Remark}

\newcommand{\Acal}{\mathcal{A}}

\newcommand{\Dcal}{\mathcal{D}}
\newcommand{\Ecal}{\mathcal{E}}

\newcommand{\Gcal}{\mathcal{G}}

\newcommand{\Lcal}{\mathcal{L}}

\newcommand{\Vcal}{\mathcal{V}}

\newcommand{\onebf}{\mathbf{1}}
\newcommand{\zerobf}{\mathbf{0}}

\newcommand{\OMIT}[1]{}

\newif\ifpdf
\ifx\pdfoutput\undefined \pdffalse \else \pdfoutput=1 \pdftrue \fi
\ifpdf
\usepackage{graphicx}
\usepackage{epstopdf}
\usepackage{graphicx}
\fi

\allowdisplaybreaks[4] 

\title{\LARGE \bf Distributed Consensus for Multiple Lagrangian Systems with Parametric Uncertainties and External Disturbances Under Directed Graphs}
\author{MEI Jie \\
\thanks{Jie Mei is with School of Mechanical Engineering and Automation, Harbin Institute of Technology, Shenzhen, Guangdong 518055, P. R.
       China.
       {Email: jmei@hit.edu.cn}.}
       \thanks{This work was supported by the National Natural Science Foundation of China (U1813220) and the Foundation Research Project of
Shenzhen (JCYJ20180306171828190, JCYJ20170413112722597).}
}

\begin{document}
%
\markboth{Accepted by IEEE Transactions on Control of Network Systems}
         {Accepted by IEEE Transactions on Control of Network Systems}

\maketitle

\begin{abstract}
In this paper, we study the leaderless consensus problem for
multiple Lagrangian systems in the presence of parametric
uncertainties and external disturbances under directed graphs. For achieving asymptotic behavior, a robust continuous term with adaptive varying gains is added to alleviate the effects of the external disturbances with unknown bounds. In the case of a fixed directed graph, by introducing an integrate term in the auxiliary variable design,
the final consensus equilibrium can be explicitly derived. We show that the agents achieve weighted average consensus, where the final equilibrium is dependent on
three factors, namely, the interactive topology, the initial positions of the agents, and the control gains
of the proposed control algorithm. In the case of switching directed graphs, a model reference adaptive consensus based algorithm is proposed such that the agents achieve leaderless consensus if  the infinite sequence of switching
graphs is uniformly jointly connected. Motivated by the fact that the relative velocity information is difficult to obtain accurately, we further propose a leaderless consensus algorithm with gain adaptation for multiple Lagrangian systems without using neighbors' velocity information. We also propose a model reference adaptive consensus based algorithm without using neighbors' velocity information for switching directed graphs. The proposed algorithms are distributed in the sense of using local information from its neighbors and using no comment control gains. Numerical simulations are performed to show the effectiveness of the proposed algorithms.
\end{abstract}


%
\IEEEpeerreviewmaketitle


\section{Introduction}
Distributed coordination of multi-agent systems has drawn a considerable attention in the last decade due to the wide applications in
unmanned aerial vehicles, sensor networks, distributed computing, as well as biology \cite{CaoYuRenChen13_TII,KnornChenMiddleton16_TCNS,QinMaShiWang17_TIE}.
In practice, a large number of physical problems can be represented by networks of agents which exchange information mutually. In such a distributed way, the whole group achieves collective behavior. 
One basic research problem is the leaderless consensus problem, where the agents achieve a common value of interest by interacting with their local neighbors. The consensus algorithm initiates the research trend in the area and triggers a lot of applications including formation\cite{OhParlAhn15_Automatica}, distributed optimization\cite{NedicOR18_PIEEE}, synchronization of biochemical networks\cite{ScardoviArcakSontag10_TAC}, and cooperative adaptive identification\cite{ChenWen14_TAC}. 

There are two important concerns in the study of consensus convergence. One is the dynamics of the agents, which includes single or double integrators\cite{RenBeardAtkins07_CSM,MeiRenChen16_TAC,LiuJiRen16_Tcyber}, general linear systems\cite{ScardoviSepulchre10_automatica,LiDuanChenHuang10_TCS1},
and nonlinear systems \cite{HouChengTan09_SMCB}. As a special case of nonlinear systems,
Lagrangian system can be used to represent a large class of
mechanical systems including robotic manipulators, autonomous vehicles, and rigid bodies \cite{KellySantibanezLoria05}. In the last decade, distributed coordination for multiple Lagrangian systems
has drawn a lot of attention\cite{ChengHouTan07,Ren09_IJC,NunoOBH11_TAC,MeiRenMa11_Automatica,MeiRenChenMa13_automatica,ZhangTangHuang18_TII,LiuJia17_IJCAS,Wang14_TAC1,Wang13_automatica,Wang17_CAC,Abdessameud18_ACC,YeAndersonYu17_IJRNC,Mei15_CAC,Mei17_ACC,
HokayemStipanovicSpong09,ChungSlotine09,MengDimaJohan14_TRO,CaiHuang16_TAC,AbdessameudTayebi17_TAC,YangFangChenJiangCao17_TAC,ChenDong17_IJRNC,LiuYeQinYu18_TSMC,
FengHuWenDixonMei18_TCNS,ChenSongGuan18_TNNLS,KlotzObuzKanDixon18_TCyber,ChopraStipanovicSpong08,GhapaniMeiRenSong16_Automatica,MengRenYou10_automaitca}.
These works include the leaderless consensus
problem\cite{ChengHouTan07,Ren09_IJC,NunoOBH11_TAC,MeiRenMa11_Automatica,MeiRenChenMa13_automatica,ZhangTangHuang18_TII,LiuJia17_IJCAS,Wang14_TAC1,Wang13_automatica,Wang17_CAC,Abdessameud18_ACC,YeAndersonYu17_IJRNC,Mei15_CAC,Mei17_ACC}, the coordinated
tracking problem with one single
leader\cite{HokayemStipanovicSpong09,ChungSlotine09,MengDimaJohan14_TRO,CaiHuang16_TAC,AbdessameudTayebi17_TAC,YangFangChenJiangCao17_TAC,ChenDong17_IJRNC,LiuYeQinYu18_TSMC,
FengHuWenDixonMei18_TCNS,ChenSongGuan18_TNNLS,KlotzObuzKanDixon18_TCyber}, the flocking with collision avoidance and/or connectivity maintenance\cite{ChopraStipanovicSpong08,GhapaniMeiRenSong16_Automatica}, and
the containment control problem with multiple
leaders\cite{MengRenYou10_automaitca,MeiRenMa11_Automatica}.
The other concern is the associated topology representing the information interaction among the agents, including undirected graphs\cite{ChengHouTan07,Ren09_IJC}, directed graphs\cite{NunoOBH11_TAC,MeiRenMa11_Automatica,MeiRenChenMa13_automatica,ZhangTangHuang18_TII,LiuJia17_IJCAS,Wang14_TAC1,Wang13_automatica,Wang17_CAC,Abdessameud18_ACC,YeAndersonYu17_IJRNC,Mei15_CAC,Mei17_ACC}, and even time-varying graphs\cite{Wang17_CAC,Abdessameud18_ACC,CaiHuang16_TAC}.

In \cite{ChengHouTan07} and \cite{Ren09_IJC}, the authors study the leaderless consensus problem for multiple Lagrangian systems under an undirected graph.
The Lyapunov based method is proposed by exploiting the symmetry property of the undirected graph. This requirement of undirected graphs might
not be practical in a realistic network, where the sensors may have different communication/sensing abilities. Instead,
it is more practical and reasonable to consider general directed graphs.
Due to the fact that the associated matrixes corresponding to directed graphs are not symmetric, it is difficult to solve the problem following the idea in the
case of undirected graphs. A common method is to introduce distributed sliding variables \cite{NunoOBH11_TAC,MeiRenMa11_Automatica,MeiRenChenMa13_automatica,ZhangTangHuang18_TII,LiuJia17_IJCAS,Wang14_TAC1,Wang13_automatica,Wang17_CAC,Abdessameud18_ACC},
inspired by the classical work of \cite{SlotineLi91}, where the control algorithms are firstly designed
for the agents such that the agents' states converge to the designed sliding surfaces. And on the sliding surfaces, the agents will achieve consensus asymptotically.
It is worthy mentioning that the sliding variable or error signal is firstly proposed in \cite{ChengHouTan07} for the leaderless consensus of multiple Lagrangian systems, however, under an undirected graph. For the consensus problem under a general directed graph, specifically, the authors in \cite{NunoOBH11_TAC} study the leaderless consensus and coordinated tracking problem with and without delays. In \cite{MeiRenMa11_Automatica}, a similar result is obtained for the leaderless consensus under a directed graph by introducing distributed sliding variables. By considering the fact that the relative velocity information is difficult to obtain accurately, the authors in \cite{MeiRenChenMa13_automatica} propose control algorithms without using relative velocity information. In \cite{ZhangTangHuang18_TII}, a time-varying sampled-date strategy is developed to realize the leaderless consensus. The consensus in the presence of external disturbances is studied in \cite{LiuJia17_IJCAS}. Note that the agents achieve consensus with a zero final velocity in \cite{Ren09_IJC,NunoOBH11_TAC,MeiRenMa11_Automatica,MeiRenChenMa13_automatica,ZhangTangHuang18_TII,LiuJia17_IJCAS}. Moreover, these existing results do not explicitly derive the final consensus equilibrium. By introducing an integral term, the scaled weighted average consensus is achieved under a directed graph in \cite{Wang14_TAC1} in the presence of communication delays. In \cite{Wang13_automatica}, the author proposes a distributed algorithm involving with integral terms for the leaderless consensus with a constant final velocity.
The stability analysis is based on frequency domain input-output method. In \cite{NunoOBH11_TAC,MeiRenMa11_Automatica,MeiRenChenMa13_automatica,ZhangTangHuang18_TII,LiuJia17_IJCAS,Wang14_TAC1,Wang13_automatica,Wang17_CAC,Abdessameud18_ACC}, adaptive controllers are proposed for the parameter
uncertainties. Recently, in \cite{YeAndersonYu17_IJRNC}, for Lagrangian systems without gravity, a distributed model-independent algorithm using only relative position and absolute velocity information is proposed to achieve leaderless consensus under a directed graph. Additional requirements on the control gains are needed.

Motivated by the previous results, we study the leaderless consensus problem for multiple Lagrangian systems in the presence of external disturbances under general directed graphs. 
A robust continuous term with adaptive varying gains is added in the control design to alleviate the effects of the external disturbances with unknown bounds, which results in an asymptotic behavior. In the case of a fixed directed graph, with the aid of an integral term in the auxiliary variable design, the final consensus equilibrium can be explicitly derived. We show that this equilibrium is dependent on three factors, namely, the interactive topology, the initial positions of the agents, and the control gains of the proposed control algorithm. Specially, only the agents who have directed paths to all the other agents are involved. Motivated by the fact that the relative velocity information is difficult to obtain accurately, we also propose a leaderless consensus algorithm with gain adaptation for multiple Lagrangian systems without using neighbors' velocity information. Lyapunov based methods are presented to show the consensus convergence, in contrast to the frequency domain input-output analysis in \cite{Wang14_TAC1}. Furthermore, the control gains for each agent are
heterogeneous and can be obtained via only local information, which makes the proposed algorithm fully distributed. In the case of switching directed graphs, model reference adaptive consensus based algorithms are proposed to solve the consensus problem under the condition that the sequence of the switching directed graphs is uniformly jointly connected. Partial of the current work has appeared in \cite{Mei15_CAC} and \cite{Mei17_ACC}. The improvements include the presence of external disturbances, the extension to a general directed graph containing a spanning tree from a strongly connected graph, and the switching directed graphs.
And the proofs are in more detail and numerical simulations are performed. Compared with the existing results, our proposed algorithms in this paper have the following advantages.
\begin{itemize}
  \item[(1)] For a fixed directed graph, weighted leaderless consensus for multiple Lagrangian systems is solved with explicitly derived final equilibrium. For switching directed graphs, leaderless consensus is solved under the wild assumption that the switching graphs are uniformly jointly connected. Both cases with and without relative velocity feedback are considered.
  \item[(2)] Asymptotical consensus convergence is achieved even in the presence of bounded external disturbances with unknown bounds, by the proposed continuous algorithms with a robust term.
  \item[(3)] The proposed algorithms are fully distributed in the sense of using local information from its neighbors and using no comment control
gains.
\end{itemize}

{\it Notations:} Let $\onebf_m$ and $\zerobf_m$ denote,
respectively, the $m\times 1$ column vector of all ones and all
zeros. Let $\zerobf_{m\times n}$ denote the $m\times n$ matrix with
all zeros and $I_m$ denote the $m\times m$ identity matrix. Let
$\mbox{diag}(z_1,\cdots,z_p)$ be the diagonal matrix with diagonal
entries  $z_1$ to $z_p$. For a time-varying vector $f(t):\re\mapsto\re^m$, it is said that
$f(t)\in\mathbb{L}_l$, $l\in[1,\infty)$, if
$\|f(t)\|_l\defeq (\int_{0}^{\infty}\|f(\tau)\|^l\mbox{d}\tau)^{\frac{1}{l}}<\infty$ and
$f(t)\in\mathbb{L}_{\infty}$ if $\|f(t)\|_\infty\defeq \sup_{t\geq 0}\|f(t)\|<\infty$. Throughout the
paper, we use $\|\cdot\|$ to denote the Euclidean norm.



\section{Background}\label{section_background}
\subsection{Euler-Lagrange System}

We consider a multi-agent systems with $n$ agents whose dynamics is represented by
the following Euler-Lagrange equation
\begin{equation}\label{eq:EL-system}
    M_i(q_i)\ddot q_i+C_i(q_i,\dot q_i)\dot
    q_i+g_i(q_i)+d_i(t)=\tau_i,~~~i=1,\cdots,n,
\end{equation}
where $q_i\in \mathbb{R}^p$ is the position vector, $\dot q_i\in\mathbb{R}^p$ is the velocity vector, $M_i(q_i)\in\mathbb{R}^{p\times p}$ is the inertia matrix, $C_i(q_i,\dot q_i)\dot q_i\in
\mathbb{R}^p$ is the Coriolis and centrifugal torques,
$g_i(q_i)$ is the gravitational torque, $d_i(t)\in\re^p$ is the external disturbance, and $\tau_i\in
\mathbb{R}^p$ is the control input on the $i$th agent. We
assume that the external disturbances $d_i(t)$, $i=1,\ldots, n$, are upper bounded by an unknown bound $d_{\max}>0$, $i.e.,$ $\|d_i(t)\|\leq d_{\max}$, $\forall i=1,\ldots,n$.

In spirt of the complexity of the equation \eqref{eq:EL-system}, which can describe the dynamics of a large class of mechanical systems, it has inherent interesting properties
that are of practice importance for the control purposes. Some of them are listed as follows, which are useful for the subsequent analysis\cite{KellySantibanezLoria05}:
\begin{itemize}
  \item[(A1)] The inertia matrix $M_i(q_i)$ is symmetric positive definite, and for any $i$, there exist positive constants $k_{\underline{m}}$, $k_{\overline{m}}$, $k_C$, and $k_{g_i}$ such that $0<k_{\underline{m}}I_p\leq M_i(q_i)\leq k_{\overline{m}}I_p$, $\|C_i(x,y)\|\leq k_C\|y\|$ for all vectors $x,y\in\mathbb{R}^p$, and $\|g_i(q_i)\|\leq k_{g_i}$.
  \item[(A2)] The matrix $\dot M_i(q_i)-2C_i(q_i,\dot q_i)$ is skew symmetric, {\it i.e.}, $x^T(\dot M_i(q_i)-2C_i(q_i,\dot q_i))x=0$, $\forall x\in\mathbb{R}^p$.
  \item[(A3)] For \eqref{eq:EL-system}, we have the following property of linear parameterization: $M_i(q_i)x+C_i(q_i,\dot q_i)y+g_i(q_i)=Y_i(q_i,\dot q_i,x,y)\Theta_i$ for all vectors $x,y\in\mathbb{R}^p$, where $Y_i(q_i,\dot q_i,x,y)$ is the regressor and $\Theta_i$ is the constant parameter vector associated with the $i$th agent. 
      In this current paper, it is assumed that the constant vector $\Theta_i$ is unknown, which represents the parametric uncertainties in the agent dynamics.
\end{itemize}


\subsection{Graph Theory}

In this paper, we use a general directed graph to model the interaction among the
$n$ agents. A directed graph of order $n$ is a pair $\mathcal {G}\defeq (\mathcal {V},\mathcal {E})$,
where $\mathcal {V}\defeq\{1,...,n\}$ is the node set and $\mathcal {E}\subseteq \mathcal {V}\times\mathcal
{V}$ is the edge set. An edge $(i,j)\in \Ecal$ denotes that agent $j$ can obtain
information from agent $i$, but not necessarily vice versa. Self edges $(i,i)$ are not allowed in this paper.
For the edge $(i,j)\in \Ecal$, node $i$ is the parent node while node $j$ is the child node. Equivalently, node
$i$ is a neighbor of node $j$. A directed path is a sequence of edges of the form $(i_1,i_2)$, $(i_2,i_3)$,
$\ldots$, in a directed graph. A directed graph is strongly
connected if there exists a directed path from every node to every other node. A directed tree is a directed graph in which every node has
exactly one parent except for one node, called the root, which has directed paths to every other node. A subgraph $\mathcal {G}^s\defeq (\mathcal {V}^s,\mathcal {E}^s)$ of $\mathcal{G}$ is a graph such that $\mathcal V^s\subseteq \mathcal V$ and $\mathcal E^s\subseteq (\mathcal V^s\times \mathcal V^s)\bigcap\mathcal E$. A directed spanning
tree $(\mathcal {V}^s,\mathcal {E}^s)$ of a directed graph $\Gcal$ is subgraph of $\Gcal$ such that $(\mathcal {V}^s,\mathcal {E}^s)$ a direct tree and $(\mathcal {V}^s=\Vcal$. A directed graph contains a spanning tree if
there exists a directed spanning tree as a subgraph of the directed
graph.

The adjacency matrix $\Acal=[a_{ij}]\in\mathbb{R}^{n\times n}$
associated with $\Gcal$ is defined as $a_{ij}>0$ if $(j,i)\in
\Ecal$, and $a_{ij}=0$ otherwise. Since self edges are not
allowed in this paper, {\it i.e.,} $a_{ii}=0$. The Laplacian matrix
$\Lcal_A=[l_{ij}]\in\mathbb{R}^{n\times n}$ associated with $\Acal$
and hence $\Gcal$ is defined as $l_{ii}=\sum_{j=1,j\neq i}^na_{ij}$
and $l_{ij}=-a_{ij}$, $i\neq j$. For a general directed graph, $\Lcal_A$
is not necessarily symmetric, which makes the consensus analysis under a
directed graph more challenging. Fortunately, we have the following nice properties on the
Laplacian matrix of a directed graph.

\begin{lemma}\cite{RenBeard08}\label{lemma:spanningtree}
Let $\Lcal_A\in\re^{n\times n}$ be the (nonsymmetric)
Laplacian matrix associate with the directed graph $\Gcal$. The following statements hold:
\begin{itemize}
  \item[1)] The matrix $\Lcal_A$ has a simple zero eigenvalue and all other eigenvalues
  have positive real parts if and only if $\Gcal$ contains a directed spanning tree;
  \item[2)]  If $\Gcal$ contains a directed spanning tree, there exists a vector $\xi\defeq
[\xi_1,\ldots,\xi_{n}]^T\in\re^{n}$ with $\sum_{i=1}^{n}\xi_i=1$ and
$\xi_i\geq 0$, $\forall i=1,\ldots,n$, such that $\xi^T \Lcal_A=0$. Furthermore, if $\Gcal$ is strongly connected, the above statement holds with $\xi_i> 0$, $\forall i=1,\ldots,n$.
\end{itemize}
\end{lemma}

\begin{lemma}\label{lemma:dir-undir}\cite{MeiRenChen16_TAC}
Let $\Lcal_A\in\re^{n\times n}$ be the (nonsymmetric)
Laplacian matrix associate with the directed graph $\Gcal$ being strongly connected. Define the matrix $B\defeq\Xi \Lcal_A+\Lcal_A^T\Xi$, where
$\Xi\defeq\mbox{diag}(\xi_1,\ldots,\xi_n)$ with $\xi_i$ defined as in Lemma \ref{lemma:spanningtree}. Then for any positive vector $\varsigma\in\re^n$, the following inequality holds,
\begin{align}\label{eq:boundvec}
a(B)\defeq \min_{\stackrel{\vartheta^T\varsigma=0}{\vartheta^T\vartheta=1} }\vartheta^TB   \vartheta>0.
\end{align}
\end{lemma}

\begin{definition}\cite{BermanPlemmons79}\label{def:M-matrix}
Let $Z_n\subset \mathbb{R}^{n\times n}$ denote the set of all square
matrices of dimension $n$ with nonpositive off-diagonal entries. A
matrix $A\in \re^{n\times n}$ is said to be a nonsingular $M$-matrix
if $A\in Z_n$ and all eigenvalues of $A$ have positive real parts.
\end{definition}


\begin{lemma}\cite{BermanPlemmons79}\label{lemma:M-matrix}
If $A\in Z_n$ is a nonsingular $M$-matrix, there exists a diagonal matrix $W \defeq \mbox{diag}(w_1, \cdots,
  w_n)$ with $w_i>0$, $\forall i=1,\ldots,n$, such that $B\defeq
  WA+A^TW>0$.
\end{lemma}

%
%
%
%
%


\section{Consensus Algorithm Design}\label{sec:wconEL}

In this section, we will propose distributed control algorithms for
each agent such that the agents achieve consensus in the presence of external disturbances and parametric uncertainties under
fixed and switching directed graphs.

\subsection{Fixed Directed Graph}

We first consider the case under a fixed directed graph.
Before presenting the main result, we would like to review the existing results on consensus of multiple Lagrangian systems under a directed graph without external disturbances. Due to the fact that the associated directed graph is non-symmetric, it is difficult to design Lyapunov functions directly following the ideas for undirected graphs. One alternative way is to introduce distributed sliding variables as follows \cite{MeiRenMa11_Automatica,NunoOBH11_TAC}
\begin{align}
  \dot q_{ri} \defeq& -\alpha \sum_{j=1}^n a_{ij}(q_i-q_j)  \label{eq:wconEL_auxivari21}\\
  s_i \defeq&~\dot q_i-\dot q_{ri}=\dot q_i+\alpha \sum_{j=1}^na_{ij}(q_i-q_j)\label{eq:wconEL_auxivari22}
\end{align}
where $\alpha$ is a positive constant, and $a_{ij}$ is the $(i,j)$th
entry of the adjacency matrix $\Acal$ associated with $\Gcal$.
And the following distributed adaptive control algorithm is proposed for each agent
\begin{subequations}\label{eq:coorEL_controlinput2}
\begin{eqnarray}
\tau_i &=& -K_{i}s_i+Y_i(q_i,\dot q_i,\ddot q_{ri},\dot q_{ri})\widehat
\Theta_i\label{eq:coorEL_controlinput21}\\
\dot{\widehat{\Theta}}_i&=&-\Lambda_iY_i^T(q_i,\dot
q_i,\ddot q_{ri},\dot q_{ri})s_i\label{eq:coorEL_adaptationlaw22}
\end{eqnarray}
\end{subequations}
where $K_i$ and $\Lambda_i$ are symmetric positive-definite
matrixes,
$\widehat{\Theta}_i$ is the estimate of $\Theta_i$, and
$Y_i(q_i,\dot q_i,{\ddot q}_{ri}, {\dot q}_{ri})$ is defined
as in (A3). The control algorithm \eqref{eq:coorEL_controlinput21} aims to
drive the agents' states to the sliding surface $s_i=0$, where \eqref{eq:coorEL_adaptationlaw22} is
used to compensate the parametric uncertainties. And on the sliding surface $s_i=0$, the agents will achieve consensus
asymptotically. The following result is presented in \cite{MeiRenMa11_Automatica} for a general fixed directed graph.

\begin{theorem}\label{thm:EL-consensus}
Using \eqref{eq:coorEL_controlinput2} for
\eqref{eq:EL-system} with $d_i(t)=0$, $\|q_i(t)-q_j(t)\|\to 0$ and $\dot
q_i(t)\to \zerobf_p$ as $t\to\infty$ if and only
if the directed graph $\Gcal$ associated with the $n$ agents contains a
directed spanning tree.
\end{theorem}

The above result shows the consensus convergence of multiple Lagrangian systems under a general directed graph in the presence of parametric uncertainties.
However, there are still some issues which need to be further studied. The first one is that the final consensus equilibrium is not explicitly derived. It is still unclear where the consensus equilibrium would be. The second is that the external disturbances are not taken into account. In this section, we aim to propose a new distributed control algorithm such that the final consensus equilibrium can be explicitly derived, which means that an asymptotic consensus convergence should be achieved even in the presence of external disturbances.
Different from \eqref{eq:wconEL_auxivari21} and \eqref{eq:wconEL_auxivari22}, we introduce the following auxiliary variables which is motivated by \cite{Wang14_TAC1}
\begin{align}
\vartheta_i=&~\dot q_i+\alpha_i\sum_{i=1}^n a_{ij}(q_i-q_j) \label{eq:wconEL_auxivari1}\\
\dot q_{ri}=&~-\alpha_i\sum_{i=1}^n a_{ij}(q_i-q_j)-\int_{0}^t\vartheta_i(\tau)\mbox{d}\tau\label{eq:wconEL_auxivari2}\\
s_i=&~\dot q_i-\dot q_{ri}=\vartheta_i+\int_{0}^t\vartheta_i(\tau)\mbox{d}\tau\label{eq:wconEL_auxivari3}
\end{align}
where $\alpha_i$ are arbitrary positive constants, $i=1,\ldots, n$. The differences compared with \eqref{eq:wconEL_auxivari21} and \eqref{eq:wconEL_auxivari22} are stated as follows. First, an integral term is introduced into the auxiliary variable design, which has benefits on deriving the final consensus equilibrium. Second, each agent is assigned its own gain $\alpha_i$. Therefore, no
common gains will be used in the control algorithm.

We propose the following control algorithm for \eqref{eq:EL-system}
\begin{subequations}\label{eq:wconEL_ctr}
\begin{eqnarray}
\tau_i\!\!\!\! &=&\!\!\!\! -K_{i}s_i+Y_i(q_i,\dot q_i,\ddot q_{ri},\dot q_{ri})\widehat
\Theta_i-\frac{\hat d_i}{\|s_i\|+\mu_i(t)}s_i\label{eq:wconEL_ctr1}\\
\dot{\widehat{\Theta}}_i\!\!\!\!&=&\!\!\!\!-\Lambda_iY_i^T(q_i,\dot
q_i,\ddot q_{ri},\dot q_{ri})s_i\label{eq:wconEL_ctr2}\\
\dot{\hat d}_i\!\!\!\!&=&\!\!\!\!\delta_i\frac{\|s_i\|^2}{\|s_i\|+\mu_i(t)}\label{eq:wconEL_ctr3}
\end{eqnarray}
\end{subequations}
where $K_i$ and $\Lambda_i$ are symmetric positive-definite
matrixes, $\delta_i$ is a positive constant, $\dot q_{ri}$ and $s_i$ are defined as in \eqref{eq:wconEL_auxivari2} and \eqref{eq:wconEL_auxivari3}, respectively,
$\widehat{\Theta}_i$ is the estimate of $\Theta_i$, $\hat d_i$ is the estimate of $d_{\max}$ with $\hat d_i(0)\geq 0$,
and $Y_i(q_i,\dot q_i,{\ddot q}_{ri}, {\dot q}_{ri})$ is defined
as in (A3). The function $\mu_{i}(t)$ is chosen such that $\mu_i(t)>0$, $\forall t\geq 0$, and $\int_0^{\infty}\mu_i(t)< \infty$. Examples include $e^{-t}$ and $\frac{1}{(t+1)^2}$. The third term in \eqref{eq:wconEL_ctr1} $-\frac{\hat d_i}{\|s_i\|+\mu_i(t)}s_i$ is used to compensate for the bounded external disturbances, which is continuous. 
Clearly, under \eqref{eq:wconEL_ctr3}, $\hat d_i(t)\geq 0$, $\forall t\geq 0$.

\begin{remark}
Note from \eqref{eq:wconEL_ctr3} that $\hat{d}_i(t)$ is monotonically increasing and sensitive to $\|s_i\|$. One alternation is to use the $\sigma-$modification, where \eqref{eq:wconEL_ctr3} can be redesigned as
\begin{align*}
  \dot{\hat d}_i=~\delta_i(\frac{\|s_i\|^2}{\|s_i\|+\mu_i(t)}-\sigma_i\hat d_i)
\end{align*}
with $\sigma_i>0$. However, only UUB (uniformly ultimately bounded) result can be obtained. One improvement
is the adaptive version of the $\sigma-$modification, where \eqref{eq:wconEL_ctr3} is redesigned as
\begin{align*}
 \dot{\hat d}_i&=~\delta_i\big[\frac{\|s_i\|^2}{\|s_i\|+\mu_i(t)}-\sigma_i(\hat d_i-\bar d_i)\big] \\
  \dot{\bar d}_i&=~\gamma_i(\hat d_i-\bar d_i)
\end{align*}
with $\delta_i,\sigma_i, \gamma_i>0$. Using this scheme, asymptotical consensus convergence can be preserved with robust and fast adaption
 in the face of high-gain learning rates. The detailed discussions can be found in \cite{YucelenHaddad13_TAC} and \cite{MeiRenChen16_TAC}.
\end{remark}

To facilitate the convergence analysis, the following $(n-1)\times n$ matrix $Q$ is introduced \cite{ScardoviArcakSontag10_TAC}
\begin{align}\label{eq:Q}
Q=\left(
    \begin{array}{ccccc}
      -1+(n-1)v & 1-v & -v & \cdots & -v \\
      -1+(n-1)v & -v & 1-v & \ddots & \vdots \\
      \vdots & \vdots & \ddots & \ddots & -v \\
      -1+(n-1)v & -v & \cdots & -v & 1-v \\
    \end{array}
  \right)
\end{align}
with $v=\frac{n-\sqrt{n}}{n(n-1)}$. The matrix $Q$ has the following properties
\begin{align}\label{Q1}
Q\onebf_n=\zerobf_{n-1},
QQ^T=I_{n-1},
Q^TQ=I_n-\frac{1}{n}\onebf_n\onebf_n^T.
\end{align}

And we have the following result.

\begin{lemma}\label{lemma:wconEL-Q}
Suppose that the directed graph $\Gcal$ contains a directed spanning tree. All the eigenvalues of $Q\Delta\Lcal_AQ^T$ have positive real parts, where $Q$ is defined in \eqref{eq:Q}, $\Lcal_A$ is the Laplacian matrix associated with $\Gcal$, and $\Delta\defeq \mbox{diag}(\alpha_1,\cdots,\alpha_n)$ with $\alpha_i$ being positive constants defined in \eqref{eq:wconEL_auxivari1}. Furthermore, for any time-varying vector $x(t)\in\re^n$, $Qx(t)\in\mathbb{L}_{\infty} \Leftrightarrow \Lcal_A x(t)\in\mathbb{L}_{\infty}$ and $\lim_{t\to\infty}\|Qx(t)\|=0 \Leftrightarrow \lim_{t\to\infty}\|\Lcal_A x(t)\|=0$.
\end{lemma}
\proof
We first introduce the following $n\times n$ augmented matrix
\begin{align}
\widehat Q\defeq \left(
               \begin{array}{c}
                 \onebf_n^T/\sqrt{n} \\
                 Q \\
               \end{array}
             \right).
\end{align}
It follows that
\begin{align*}
\widehat Q \widehat Q^T=& \left(
               \begin{array}{c}
                 \onebf_n^T/\sqrt{n} \\
                 Q \\
               \end{array}
             \right)\left(
                      \begin{array}{cc}
                        \onebf_n/\sqrt{n} & Q^T \\
                      \end{array}
                    \right)\\
                    =&\left(
                              \begin{array}{cc}
                                1 & \onebf_n^TQ^T/\sqrt{n} \\
                                Q\onebf_n/\sqrt{n} & QQ^T \\
                              \end{array}
                            \right),
\end{align*}
and
\begin{align*}
\widehat Q^T \widehat Q=& \left(
                      \begin{array}{cc}
                        \onebf_n/\sqrt{n} & Q^T \\
                      \end{array}
                    \right)\left(
               \begin{array}{c}
                 \onebf_n^T/\sqrt{n} \\
                 Q \\
               \end{array}
             \right)\\
             =&~\frac{1}{n}\onebf_n\onebf_n^T+Q^TQ.
\end{align*}
Using \eqref{Q1}, we can get
$\widehat Q\widehat Q^T=\widehat Q^T\widehat Q=I_n$,
and thus $\widehat Q$ is unitary. Note that
\begin{align}
\widehat Q \Delta\Lcal_A \widehat Q^T=&~\left(
               \begin{array}{c}
                 \onebf_n^T/\sqrt{n} \\
                 Q \\
               \end{array}
             \right)\Delta\Lcal_A\left(
                      \begin{array}{cc}
                        \onebf_n/\sqrt{n} & Q^T \\
                      \end{array}
                    \right)\notag\\
                    =&~\left(
                         \begin{array}{cc}
                           \frac{\onebf_n^T\Delta\Lcal_A\onebf_n}{n} & \frac{\onebf_n^T\Delta\Lcal_AQ^T}{\sqrt{n}} \\
                           \frac{Q\Delta\Lcal_A\onebf_n}{\sqrt{n}} & Q\Delta\Lcal_AQ^T \\
                         \end{array}
                       \right)\notag\\
                       =&\left(
                         \begin{array}{cc}
                           0 & \frac{\onebf_n^T\Delta\Lcal_AQ^T}{\sqrt{n}} \\
                           \zerobf_{n-1} & Q\Delta\Lcal_AQ^T \\
                         \end{array}
                       \right). \label{eq:wconEL-Qhat}
\end{align}
The above equality as well as the fact that $\widehat Q$ is unitary imply that the $n$ eigenvalues of $\Delta\Lcal_A$ contains zero and the $n-1$ eigenvalues of $Q\Delta\Lcal_AQ^T$.
On the other hand, note that $\Delta\Lcal_A$ can be viewed as the Laplacian matrix associated with the graph $\Gcal$, with the adjacency matrix $\Acal=[a_{ij}]\in\mathbb{R}^{n\times n}$
being replaced by $\Delta\Acal=[\alpha_ia_{ij}]$. Therefore,
if $\Gcal$ contains a directed spanning tree, we can get from Lemma \ref{lemma:spanningtree} that the matrix $\Delta\Lcal_A$ has one single zero eigenvalue and all other eigenvalues
have positive real parts. We then can get from \eqref{eq:wconEL-Qhat} that all the eigenvalues of $Q\Delta\Lcal_AQ^T$ have positive real parts.

For any time-varying vector $x$, if $Qx\in\mathbb{L}_\infty$, $\Lcal_AQ^TQx\in\mathbb{L}_\infty$ since multiplying a bounded matrix does not change the boundedness.
From \eqref{Q1},
\begin{align}\label{eq:Q4}
\Lcal_AQ^TQx=\Lcal_A(I_n-\frac{1}{n}\onebf_n^T\onebf_n)x=\Lcal_Ax.
\end{align}
We then can conclude that $\Lcal_A x\in\mathbb{L}_\infty$. On the other hand, if $\Lcal_Ax\in\mathbb{L}_\infty$, from \eqref{eq:Q4}, $\Lcal_AQ^TQx\in\mathbb{L}_\infty$.
And $Q\Lcal_AQ^TQx\in\mathbb{L}_\infty$. Note that the matrix $Q\Lcal_AQ^T$ is nonsingular under the condition that $\Gcal$ contains a directed spanning tree. We then can get
$Qx\in\mathbb{L}_\infty$. Following the same process, one can easily get that $\lim_{t\to\infty}\|Qx(t)\|=0 \Leftrightarrow \lim_{t\to\infty}\|\Lcal_A x(t)\|=0$.
\endproof

\begin{remark}
From Lemma \ref{lemma:wconEL-Q}, the $n-1$ eigenvalues of $Q\Delta\Lcal_AQ^T\in\re^{(n-1)\times (n-1)}$ are exactly the $n-1$ ones with positive real parts of $\Delta\Lcal_A$. Actually, the transformation matrix $Q$ is introduced to
convert the consensus problem to a stabilization problem.
\end{remark}

We have the following main result under a fixed directed graph.

\begin{theorem}\label{thm:wconEL}
Suppose that the directed graph $\Gcal$ contains a directed spanning tree. Using \eqref{eq:wconEL_ctr} for
\eqref{eq:EL-system}, $\|q_i(t)-q_j(t)\|\to 0$ and $\dot
q_i(t)\to \zerobf_p$ as $t\to\infty$. In particular, $\lim_{t\to\infty} q_i(t)=\sum_{i=1}^n \frac{\xi_i}{\alpha_i} q_i(0)/\sum_{i=1}^n \frac{\xi_i}{\alpha_i}$, $i=1,\ldots, n$, where $\xi_i$ is defined as in Lemma \ref{lemma:spanningtree}, and $\alpha_i$ is defined in \eqref{eq:wconEL_auxivari1}.
\end{theorem}
\proof  Using \eqref{eq:wconEL_ctr} for \eqref{eq:EL-system}, we have the following closed-loop system
\begin{align}
    M_i(q_i)\dot{s}_i=&-C_i(q_i,\dot q_i)s_i-K_i s_i-Y_i(q_i,\dot q_i,{\ddot q}_{ri},
{\dot q}_{ri})\widetilde{\Theta}_i \notag\\
&-d_i(t)-\frac{\hat d_i}{\|s_i\|+\mu_i(t)}s_i \label{eq:wconEL_closeloop1}
\end{align}
where $\widetilde{\Theta}_i\defeq \Theta_i-\widehat\Theta_i$.

Considering the following Lyapunov function candidate
\begin{align}\label{eq:wconEL-Lya}
V=&\frac{1}{2}\sum_{i=1}^n s_i^TM_i(q_i)s_i+\frac{1}{2}\sum_{i=1}^n\widetilde\Theta_i^T\Lambda_i^{-1}\widetilde\Theta_i\notag\\
&+\sum_{i=1}^n\frac{1}{2\delta_i}(\hat d_i-d_{\max})^2.
\end{align}
where $d_{\max}$ is the unknown upper bound for the external disturbances. The derivative of $V(t)$ along \eqref{eq:wconEL_closeloop1} can be computed as follows
\begin{align}\label{eq:wconEL_Vdot}
\dot V=& \sum_{i=1}^n \big[s_i^TM_i(q_i)\dot s_i+\frac{1}{2}s_i^T\dot M_i(q_i) s_i +\widetilde\Theta_i^T\Lambda_i^{-1}\dot{\widetilde\Theta}_i\notag\\
&+\frac{1}{\delta_i}(\hat d_i-d_{\max})\dot{\hat d}_i\big]\notag\\
\leq & \sum_{i=1}^n\big[-s_i^TK_is_i+\|s_i\|d_{\max}-\frac{d_{\max}\|s_i\|^2}{\|s_i\|+\mu_i(t)}\big]\notag\\
=& \sum_{i=1}^n\big[-s_i^TK_is_i+\frac{d_{\max}\|s_i\|\mu_i(t)}{\|s_i\|+\mu_i(t)}\big]\notag\\
\leq& \sum_{i=1}^n\big[-s_i^TK_is_i+d_{\max}\mu_i(t)\big]
\end{align}
where we have used the fact that $\mu_i(t)>0$ to obtain the last inequality.
Integrating both sides of \eqref{eq:wconEL_Vdot}, we have
\begin{align*}
&V(t)-V(0) \\
\leq&~ \sum_{i=1}^n\big[-\int_{0}^ts_i^T(\tau)K_is_i(\tau)\mbox{d}\tau+d_{\max}\int_{0}^t\mu_i(\tau)\mbox{d}\tau\big]
\end{align*}
which can be rewritten as
\begin{align*}
&V(t)+\sum_{i=1}^n\int_{0}^ts_i^T(\tau)K_is_i(\tau)\mbox{d}\tau \\
\leq&~ V(0)+d_{\max}\sum_{i=1}^n\int_{0}^t\mu_i(\tau)\mbox{d}\tau
\end{align*}
Because $\int_{0}^t\mu_i(\tau)\mbox{d}\tau<\infty$ and $K_i$, $i=1,\ldots,n$, are symmetric positive definite, we can get $s_i\in\mathbb{L}_2$ and $V(t)\in\mathbb{L}_{\infty}$, which implies that $s_i, \widetilde\Theta_i, \hat d_i\in \mathbb{L}_{\infty}$. For the system \eqref{eq:wconEL_auxivari3}, taking $s_i$ as the input and $\int_0^t\vartheta_i(\tau)\mbox{d}\tau$ the state, the system \eqref{eq:wconEL_auxivari3} is input-to-state
stable. Then it follows from $s_i\in\mathbb{L}_{\infty}$ that $\vartheta_i, \int_0^t\vartheta_i(\tau)\mbox{d}\tau\in\mathbb{L}_{\infty}$.

Let $\vartheta$, $q$, and $s$ be the column stack vectors of, respectively, $\vartheta_i$, $q_i$, and $s_i$, $i=1,\ldots, n$.
Then \eqref{eq:wconEL_auxivari1} can be written in the following vector form
\begin{align}\label{eq:wconEL_varvec}
\vartheta=\dot q +(\Delta\Lcal_A\otimes I_p) q.
\end{align}
where $\Delta$ is defined in Lemma \ref{lemma:wconEL-Q}.
Define the following vectors
\begin{align*}
\hat\vartheta\defeq (Q\otimes I_p)\vartheta, ~~\hat q\defeq (Q\otimes I_p)q,~~\hat s \defeq (Q\otimes I_p)s
\end{align*}
with $Q$ being defined in \eqref{eq:Q}. Since $\vartheta, s\in\mathbb{L}_{\infty}$, we can easily get $\hat\vartheta, \hat s\in\mathbb{L}_{\infty}$.
Note that $\Lcal_A\onebf_n=\zerobf$.
Multiplying both sides of \eqref{eq:wconEL_varvec} by $Q\otimes I_p$, we obtain
\begin{align}\label{eq:wconEL_varhatvec}
\hat \vartheta = &~ \dot{\hat q} + (Q\Delta\Lcal_A\otimes I_p) q\notag \\
               = &~ \dot{\hat q} + (Q\Delta\Lcal_AQ^TQ\otimes I_p) q \notag\\
               = &~ \dot{\hat q} + (Q\Delta\Lcal_AQ^T\otimes I_p)\hat q
\end{align}
where we have used the fact that
$\Lcal_AQ^TQ=\Lcal_A(I_n-\frac{1}{n}\onebf_n\onebf_n^T)=\Lcal_A$ from \eqref{Q1} to obtain the second equality.
Also note that $\mbox{rank}(Q)=\mbox{rank}(QQ^T)=n-1$ and thus the null space of $Q$ is $\{a\onebf_n: a\in \re\}$. In other words, $Qx=\zerobf$ if and only if $x=a\onebf_n$ for some $a\in\re$, which means that the agents achieve consensus ($q=b\onebf_n$ for some $b\in\re^p$) if and only if $\hat q=\zerobf_{p(n-1)}$. Therefore, the consensus problem for \eqref{eq:EL-system} using \eqref{eq:wconEL_ctr} is converted into the stability problem of system \eqref{eq:wconEL_varhatvec}. One key is the properties of the matrix $Q\Delta\Lcal_AQ^T$, especially the distribution of its eigenvalues.

Since $\Gcal$ contains a directed spanning tree, we can get from Lemma \ref{lemma:wconEL-Q}
that all the eigenvalues of $Q\Delta\Lcal_AQ^T$ have positive real parts. For the system \eqref{eq:wconEL_varhatvec}, taking $\hat\vartheta$ as the input and $\hat q$ the state, 
the system \eqref{eq:wconEL_varhatvec} is input-to-state
stable. Then from 
$\hat\vartheta\in\mathbb{L}_{\infty}$, we can obtain that $\hat q\in\mathbb{L}_{\infty}$ and thus $\dot {\hat q} \in\mathbb{L}_{\infty}$ from \eqref{eq:wconEL_varhatvec} and $(\Lcal_A\otimes I_p) q\in \mathbb{L}_{\infty}$. From \eqref{eq:wconEL_auxivari1} and $\hat\vartheta\in\mathbb{L}_{\infty}$, we can get $\dot q\in\mathbb{L}_{\infty}$. From \eqref{eq:wconEL_auxivari2} and $\int_0^t\vartheta_i(\tau)\mbox{d}\tau\in\mathbb{L}_{\infty}$, we have $\dot q_{ri}\in\mathbb{L}_\infty$. Differentiating both sides of \eqref{eq:wconEL_auxivari2}, we can conclude from $\dot q, \vartheta \in\mathbb{L}_{\infty}$ that $\ddot q_{ri}\in\mathbb{L}_\infty$. Up to now, we obtain $s_i, \dot q_i, \dot q_{ri}, \ddot q_{ri}, \widetilde\Theta_i,\hat d_i\in\mathbb{L}_{\infty}$. We then can get from (A1) and \eqref{eq:wconEL_closeloop1} that $\dot s_i\in\mathbb{L}_{\infty}$. Therefore, we have $s_i\in\mathbb{L}_2\bigcap\mathbb{L}_{\infty}$ and $\dot s_i\in\mathbb{L}_{\infty}$. Then from Barbalat's Lemma, we can conclude that $\lim_{t\to\infty} \|s_i(t)\|=0$, $i=1,\ldots, n$.

Note that the system \eqref{eq:wconEL_auxivari3} is input-to-state
stable with respect to the input $s_i$ and the state $\int_0^t\vartheta_i(\tau)\mbox{d}\tau$. Since $\lim_{t\to\infty} \|s(t)\|=0$, 
 $\lim_{t\to\infty} \|\int_0^t\vartheta_i(\tau)\mbox{d}\tau\|=0$ and $\lim_{t\to\infty} \|\vartheta_i(t)\|=0$, $i=1,\ldots, n$. Clearly, $\lim_{t\to\infty} \|\hat \vartheta(t)\|=0$.
From \eqref{eq:wconEL_varhatvec}, we can obtain that $\lim_{t\to\infty} \|\hat q\|=0$.
From Lemma \ref{lemma:wconEL-Q}, we can conclude that $\lim_{t\to\infty}\|(\Lcal_A\otimes I_p)q\|=0$, {\it i.e.}, $\lim_{t\to\infty}\|q_i(t)-q_j(t)\|=0$. Since $\lim_{t\to\infty} \|\vartheta(t)\|=0$, we can get from \eqref{eq:wconEL_varvec} that $\lim_{t\to\infty} \|\dot{q}\|=0$.

In the remainder of this proof, we derive the final consensus equilibrium.  Since $\lim_{t\to\infty}\|q_i(t)-q_j(t)\|=0$, $\forall i,j=1,\ldots,n$, there exists a function $q_{\infty}(t)$ such that $\lim_{t\to\infty} (q_i(t)-q_\infty(t))=\zerobf_p$, $\forall i=1,\ldots,n$. Note that $q_\infty(t)$ can be either a constant vector or a time-varying function. It is not clear that $q_i(t)$ converges currently. Multiplying both sides of \eqref{eq:wconEL_varvec} by $(\xi^T\Delta^{-1})\otimes I_p$, where $\xi$ is 
defined in Lemma \ref{lemma:spanningtree}. And we have
\begin{align*}
[(\xi^T\Delta^{-1})\otimes I_p]\vartheta = & [(\xi^T\Delta^{-1})\otimes I_p]\dot q +[(\xi^T\Delta^{-1}\Delta\Lcal_A)\otimes I_p]q \\
=& [(\xi^T\Delta^{-1})\otimes I_p]\dot q
\end{align*}
where we have used the fact $\xi^T\Lcal_A=\zerobf_n$ to obtain the last equality. We then have
\begin{align}\label{eq:wconEL_99}
\sum_{i=1}^n \frac{\xi_i}{\alpha_i}\vartheta_i =  \sum_{i=1}^n \frac{\xi_i}{\alpha_i}\dot q_i.
\end{align}
Integrating both sides of \eqref{eq:wconEL_99} from $0$ to $t$, we have
\begin{align}\label{eq:wconEL-101}
&\sum_{i=1}^n \frac{\xi_i}{\alpha_i}\int_0^{t}\vartheta_i(\tau)\mbox{d}\tau \notag\\
 =&~\sum_{i=1}^n \frac{\xi_i}{\alpha_i}\int_0^{t}\dot q_i(\tau)\mbox{d}\tau\notag\\
=&~\sum_{i=1}^n \frac{\xi_i}{\alpha_i} \big(q_i(t)-q_i(0)\big)\notag\\
=&~\sum_{i=1}^n \frac{\xi_i}{\alpha_i} q_\infty(t)-\sum_{i=1}^n \frac{\xi_i}{\alpha_i} q_i(0) +\sum_{i=1}^n \frac{\xi_i}{\alpha_i} \big(q_i(t)-q_{\infty}(t)\big).
\end{align}
Noting that $\lim_{t\to\infty}\big(q_i(t)-q_{\infty}(t)\big)=\zerobf_p$, $\forall \varepsilon>0, \exists T_1>0$, such that $\|q_i(t)-q_{\infty}(t)\|<\frac{\varepsilon}{2}$, $\forall t>T_1$. Similarly, since $\lim_{t\to\infty} \int_0^t\vartheta_i(\tau)\mbox{d}\tau=\zerobf_p$, $\forall \varepsilon>0, \exists T_2>0$, such that $\| \int_0^t\vartheta_i(\tau)\mbox{d}\tau\|<\frac{\varepsilon}{2}$, $\forall t>T_2$. From \eqref{eq:wconEL-101}, when $t>\max\{T_1,T_2\}$
\begin{align*}
  &\|q_\infty(t)-\sum_{i=1}^n \frac{\xi_i}{\alpha_i} q_i(0)/\sum_{i=1}^n \frac{\xi_i}{\alpha_i}\| \\
  = &~ \|\sum_{i=1}^n \frac{\xi_i}{\alpha_i}\big(\int_0^{t}\vartheta_i(\tau)\mbox{d}\tau-q_i(t)+q_{\infty}(t)\big)/\sum_{i=1}^n \frac{\xi_i}{\alpha_i}\| \\
  \leq &~ \sum_{i=1}^n \frac{\xi_i}{\alpha_i}\big(\|\int_0^{t}\vartheta_i(\tau)\mbox{d}\tau\|+\|q_i(t)-q_{\infty}(t)\|\big)/\sum_{i=1}^n \frac{\xi_i}{\alpha_i}\\
  \leq &~\varepsilon
\end{align*}
which implies that the limit $\lim_{t\to\infty}q_\infty(t)$ exists. 
Therefore, we can get that the limit $\lim_{t\to\infty}q_i(t)$ also exists and
\begin{align}\label{eq:wconEL-100}
\lim_{t\to\infty} q_i(t)= \lim_{t\to\infty}q_\infty(t)=\sum_{i=1}^n \frac{\xi_i}{\alpha_i} q_i(0)/\sum_{i=1}^n \frac{\xi_i}{\alpha_i}.
\end{align}

Eq. \eqref{eq:wconEL-100} shows that the positions of the agents will converge to a stationary point, which is the weighted average of the initial positions.
\endproof

By introducing a robust term with adaptive varying gains in the control design, the asymptotic consensus has
been achieved even in the presence of external disturbances. And the final consensus equilibrium has been derived with the help of integral terms in \eqref{eq:wconEL_auxivari2}, which is dependent on the initial states of the agents, the interactive topology, and the control gains of the proposed control algorithm.

We next show that the final consensus equilibrium depends on the agents which are the roots of the graph. Without loss of generality, we can assume that the Laplacian matrix
$\Lcal_A$ has the following form
\begin{align}\label{eq:coorEL-PFform1}
\Lcal_A=\left(\begin{array}{cc} L_{11} & 0 \\
L_{21} & L_{22}
\end{array}\right)
\end{align}
where $L_{ii}\in\re^{r_i\times r_i}$ , and $r_1+r_2=n$. Under the condition that $\Gcal$ contains a spanning tree, the agents associated with $L_{11}$ are all the roots in the graph, which implies that the directed subgraph
associated with $L_{11}\in\re^{r_1\times r_1}$ is strongly
connected. If the Laplacian matrix does not have the form of \eqref{eq:coorEL-PFform1}, one can always rearrange the order of the
agents to make the new Laplacian matrix have the form of \eqref{eq:coorEL-PFform1}.
Since $L_{11}$ is strongly connected, we can get from Lemma
\ref{lemma:spanningtree} that there exists a vector $\bar\xi\defeq
[\xi_1,\ldots,\xi_{r_1}]^T\in\re^{r_1}$ with
$\sum_{i=1}^{r_1}\xi_i=1$ and $\xi_i> 0$, $\forall i=1,\ldots,r_1$,
such that $\bar\xi^T L_{11}=0$. Define $\xi\defeq [\xi_1, \ldots, \xi_{r_1}, 0, \ldots, 0]^T$. We have $\xi^T\Lcal_A=0$.
Then the final consensus equilibrium becomes $q(\infty)=\sum_{i=1}^{r_1} \frac{\xi_i}{\alpha_i} q_i(0)/\sum_{i=1}^{r_1} \frac{\xi_i}{\alpha_i}$, {\it i.e.}, only the roots of the graph are involved.

\subsection{Switching Directed Graphs}
In practice, the communication or sensing topology among the agents may switch due to vehicle motion or communication dropouts.
We thus get a time-varying directed graph $\Gcal(t)\defeq (\Vcal,\Ecal(t))$, where the node set $\Vcal$ is the same as the fixed one and
the edge set $\Ecal(t)$ is time-varying. The adjacency matrix $\Acal(t)=[a_{ij}(t)]\in\mathbb{R}^{n\times n}$
associated with $\Gcal(t)$ is piecewise continuous, and $a_{ij}(t)\in[\underline{a},\bar{a}]$, where $0<\underline{a}<\bar a$, if $(j,i)\in
\Ecal$, and $a_{ij}=0$ otherwise. Let $t_0, t_1,\ldots$ be the time sequence corresponding to the times at which $\Acal(t)$ switches, where it is assumed that
$t_i-t_{i-1}\geq t_D$, $\forall i=1,2,\ldots$ with $t_D$ a positive constant. An infinite sequence of switching graphs $\Gcal(t_i)$, $t=0,1,\ldots$ is called to be
uniformly jointly connected if there exists an infinite sequence of contiguous, nonempty, uniformly bounded time-intervals $[t_{i_j},t_{i_{j+1}}]$, $j=1,2,\ldots,$ starting
at $t_{i_1}=t_0$, satisfying that the union of the directed graphs across each such interval contains a directed spanning tree.

For the consensus of multiple single integrators under switching directed graphs,
\begin{align}\label{eq:wconEL-1or}
  \dot x_i  = -\alpha_i\sum_{j=1}^{n}a_{ij}(t)(x_i-x_j)
\end{align}
where $x_i\in\re$, $i=1,\ldots,n$, the following well-known result holds \cite{RenBeard08}.

\begin{lemma}\label{lemma:con-1or}
If the infinite sequence of switching graphs $\Gcal(t_i)$, $t=0,1,\ldots$ is uniformly jointly connected, the closed-loop system of \eqref{eq:wconEL-1or} is uniformly stable.
\end{lemma}

In \eqref{eq:wconEL_ctr}, $\ddot q_{ri}$ is used in the control design, which may cause a problem when $a_{ij}$ is time-varying.
Therefore, the auxiliary variables defined in \eqref{eq:wconEL_auxivari1}-\eqref{eq:wconEL_auxivari3} fails to address the consensus problem under switching directed graphs.
We next present a new approach motivated by the model reference adaptive consensus scheme proposed in our recent work \cite{Mei18_CDC}. Define $w_i\defeq \dot q_i+q_i$. Then \eqref{eq:EL-system} can be written as follows
\begin{align}\label{eq:WconEL-clo-w}
M_i(q_i)\dot w_i+C_i(q_i,\dot q_i)w_i=&~\tau_i-d_i+M_i(q_i)\dot q_i\notag\\
&+C_i(q_i,\dot q_i)q_i-g_i(q_i)
\end{align}
Note that \eqref{eq:WconEL-clo-w} is a first order nonlinear system. Motivated by the model reference adaptive consensus scheme in \cite{Mei18_CDC}, we propose the following reference model for \eqref{eq:WconEL-clo-w}
\begin{align}\label{eq:WconEL-z}
  \dot z_i & =-\alpha_i\sum_{j=1}^n a_{ij}(t)(w_i-w_j) \notag \\
  &= -\alpha_i\sum_{j=1}^n a_{ij}(t)(q_i-q_j+\dot q_i-\dot q_j)
\end{align}
where $\alpha_i$, $i=1,\ldots, n$, are positive constants. Clearly, both relative position and velocity measurements are used to generate the reference state $z_i$.
Define the tracking error $e_i\defeq w_i-z_i$. We have the following form of \eqref{eq:WconEL-clo-w}
\begin{align*}
&M_i(q_i)\dot e_i+C_i(q_i,\dot q_i)e_i  \\
=&~\tau_i-d_i-M_i(q_i)(\dot z_i-\dot q_i)-g_i(q_i)-C_i(q_i,\dot q_i)(z_i-q_i)\notag\\
=&~\tau_i-d_i-Y_i(q_i,\dot q_i,\dot z_i-\dot q_i,z_i-q_i)\widehat{\Theta}_i
\end{align*}

To make $e_i(t)\to \zerobf_p$ as $t\to\infty$,
we propose the following control algorithm
\begin{subequations}\label{eq:wconEL_ctr-sw}
\begin{eqnarray}
\tau_i &=& -K_{i}e_i+Y_i(q_i,\dot q_i,\dot q_i-\dot z_i,q_i-z_i)\widehat
\Theta_i\notag\\
&&-\frac{\hat d_i}{\|e_i\|+\mu_i(t)}e_i\label{eq:wconEL_ctr1-sw}\\
\dot{\widehat{\Theta}}_i&=&-\Lambda_iY_i(q_i,\dot q_i,\dot q_i-\dot z_i,q_i-z_i)e_i\label{eq:wconEL_ctr2-sw}\\
\dot{\hat d}_i&=&\delta_i\frac{\|e_i\|^2}{\|e_i\|+\mu_i(t)}\label{eq:wconEL_ctr3-sw}
\end{eqnarray}
\end{subequations}
where $K_i, \Lambda_i, \delta_i$, and $\mu_i(t)$ are defined the same as in \eqref{eq:wconEL_ctr}. We have the following result under switching directed graphs.
\begin{theorem}\label{thm:wconEL-sw}
Suppose that the infinite sequence of switching graphs $\Gcal(t_i)$, $t=0,1,\ldots$ is uniformly jointly connected. Using \eqref{eq:wconEL_ctr-sw} for
\eqref{eq:EL-system}, $\|q_i(t)-q_j(t)\|\to 0$ and $\|\dot q_i(t)\|\to 0$ as $t\to\infty$.
\end{theorem}
\proof  Using \eqref{eq:wconEL_ctr-sw} for \eqref{eq:EL-system}, we have the following closed-loop system
\begin{align}\label{eq:wconEL_closeloop1-sw}
    M_i(q_i)\dot{e}_i=&-C_i(q_i,\dot q_i)e_i-Y_i(q_i,\dot q_i,\dot q_i-\dot z_i,q_i-z_i)\widetilde{\Theta}_i \notag\\
&-K_i s_i-d_i(t)-\frac{\hat d_i}{\|e_i\|+\mu_i(t)}e_i.
\end{align}

By considering the following Lyapunov function candidate
\begin{align*}
V=&~\frac{1}{2}\sum_{i=1}^n e_i^TM_i(q_i)e_i+\frac{1}{2}\sum_{i=1}^n\widetilde\Theta_i^T\Lambda_i^{-1}\widetilde\Theta_i\notag\\
&+\sum_{i=1}^n\frac{1}{2\delta_i}(\hat d_i-d_{\max}-1)^2
\end{align*}
and following the same steps in the proof of Theorem \ref{thm:wconEL}, we can obtain
\begin{align}\label{eq:wconEL-Vdot-sw}
  \dot V\leq& \sum_{i=1}^n\big[-e_i^TK_ie_i+d_{\max}\mu_i(t)-\frac{\|e_i\|^2}{\|e_i\|+\mu_i(t)}\big]
\end{align}
Integrating both sides of \eqref{eq:wconEL-Vdot-sw}, we can get
that $e_i,\widetilde\Theta_i,\hat d_i\in\mathbb{L}_\infty$, $e_i\in\mathbb{L}_2$, and $\frac{\|e_i\|^2}{\|e_i\|+\mu_i(t)}\in\mathbb{L}_1$.
Note that
\begin{align}\label{eq:wconEL-e1-sw}
  \int_{0}^{t}\|e_i(\tau)\|\mbox{d}\tau &\! =\!\int_{0}^{t}\frac{\|e_i(\tau)\|^2}{\|e_i(\tau)\|\!+\!\mu_i(\tau)}\mbox{d}\tau \!+\!\int_{0}^{t}\frac{\|e_i(\tau)\|\mu_i(\tau)}{\|e_i(\tau)\|\!+\!\mu_i(\tau)}\mbox{d}\tau\notag\\
  &\leq \int_{0}^{t}\frac{\|e_i(\tau)\|^2}{\|e_i(\tau)\|+\mu_i(\tau)}\mbox{d}\tau +\int_{0}^{t}\mu_i(\tau)\mbox{d}\tau.
\end{align}
Since $\frac{\|e_i\|^2}{\|e_i\|+\mu_i(t)}\in\mathbb{L}_1$ and $\mu_i(t)\in\mathbb{L}_1$, we can get $e_i(t)\in\mathbb{L}_1$.

Let $z$ and $e$ be, respectively, the stack vectors of $z_i$ and $e_i$, $i=1,\ldots,n$.
From the definition of $e_i$, \eqref{eq:WconEL-z} can be written in the following vector form
\begin{align}\label{eq:wconEL-z-vec}
  \dot z & =-(\Delta\Lcal_A(t)\otimes I_p)z-(\Delta\Lcal_A(t)\otimes I_p)e.
\end{align}
Let $\Phi(t,0)$ be the transition matrix for $-\Delta\Lcal_A(t)\otimes I_p$. If the sequence of switching graphs $\Gcal(t_i)$, $i=0,1,\ldots$ is uniformly jointly connected, from Lemma \ref{lemma:con-1or}, the system $\dot{z} = -(\Delta\Lcal_A(t)\otimes I_p)z$ is uniformly stable, which implies that
$\|\Phi(t,\tau)\|\leq \gamma$, for some positive constant $\gamma$ \cite{Rugh96}. Then the solution of \eqref{eq:wconEL-z-vec} is
\begin{align}\label{eq:WconEL-z-solution}
   z(t) & = ~\Phi(t,0)z(0)-\int_{0}^{t}\Phi(t,\tau)(\Delta\Lcal_A(\tau)\otimes I_p)e(\tau)\mbox{d}\tau.
\end{align}
Since $\alpha_i$ are constants and $a_{ij}(t)$ are bounded, there exists a positive constant $c>0$ such that $\|\Delta\Lcal_A(\tau)\otimes I_p\|\leq c$. Combing with $\|\Phi(t,\tau)\|\leq \gamma$, we have
\begin{align}\label{eq:WconEL-z-bound}
  \| z(t)\|  \leq &~\|\Phi(t,0)z(0)\|+\int_{0}^{t}\|\Phi(t,\tau)\|\|\Delta\Lcal_A(\tau)\otimes I_p\|\|e(\tau)\|\mbox{d}\tau \notag\\
    \leq&~ \gamma\|z(0)\| + \gamma c\int_{0}^{t}\|e(\tau)\|\mbox{d}\tau<\infty
\end{align}
which implies that $z(t)\in\mathbb{L}_\infty$. Since $w_i=e_i+z_i$ and $e_i\in\mathbb{L}_\infty$, $w_i\in\mathbb{L}_\infty$. Note that the system $\dot q_i=-q_i+w_i$
is input-to-state stable with respect to the input $w_i$ and the state $q_i$, we can conclude that $q_i,\dot q_i\in\mathbb{L}_\infty$. From \eqref{eq:WconEL-z}, we have $\dot z_i\in\mathbb{L}_\infty$. From (A1) and \eqref{eq:wconEL_closeloop1-sw}, we can get $\dot e_i\in\mathbb{L}_\infty$. Remembering that $e_i\in\mathbb{L}_\infty\bigcap\mathbb{L}_2$, we can get from Barbalat's Lemma that $\lim_{t\to\infty}e_i(t)=\zerobf_p$, $i=1,\ldots,n$.

Define $\hat z= (Q\otimes I_p)z$, $\hat e=(Q\otimes I_p)e$,
$\hat w=(Q\otimes I_p)w$, and $\hat q=(Q\otimes I_p)q$. Clearly, $\hat e\in\mathbb{L}_\infty\bigcap\mathbb{L}_2$.
Following \eqref{eq:wconEL_varhatvec}, \eqref{eq:wconEL-z-vec} can be rewritten as
\begin{align}\label{eq:wcoEL-zclo}
  \dot{\hat z} = -(Q\Delta\Lcal_A(t)Q^T\otimes I_p)\hat z -(Q\Delta\Lcal_A(t)\otimes I_p)e
\end{align}
Since the system $\dot{z} = -(\Delta\Lcal_A(t)\otimes I_p)z$ is uniformly stable, from the properties of $Q$, the system $\dot{\hat z} = -(Q\Delta\Lcal_A(t)Q^T\otimes I_p)\hat z$ is uniformly exponentially stable, which implies that the system \eqref{eq:wcoEL-zclo} is input-to-state stable with respect to the input $(Q\Delta\Lcal_A(t)\otimes I_p)e$ and state $\hat z$. Note that $a_{ij}{(t)}$ is bounded. We can obtain that $\lim_{t\to\infty}\|\hat z(t)\|=0$. Following the same steps in the proof of Theorem \ref{thm:wconEL}, we can conclude the result.
\endproof

\section{Consensus Algorithm Without Relative Velocity Feedback}\label{sec:wconEL}

In practice, for second-order systems, relative position information among the agents can be measured by sonar or visual devices. Generally, relative velocity measurements are more difficult to obtain than relative position measurements. One way is to use differentiators from the relative position measurements. However, the differentiators are difficult to implement and extremely sensitive to errors and noises. Another way is to communicate the velocity measurements if each agent can measure its own absolute velocity, which will require the systems to be equipped with the communication capability and raise the communication burden. And in the control algorithm design for multi-agent systems, using less relative information is always welcome. Therefore, in this section, we will propose control algorithm without using neighbors' velocity measurements under both fixed and switching directed graphs.

\subsection{Fixed Directed Graph}
For a fixed directed graph, we propose a control algorithm
without using neighbors' velocity measurements motivated by \cite{MeiRenChenMa13_automatica}
\begin{subequations}\label{eq:wconELwoRV_ctr}
\begin{eqnarray}
\tau_i &=&-\hat k_{i}(t)s_i+Y_i(q_i,\dot q_i,\zerobf_p,\dot q_{ri})\widehat
\Theta_i\notag\\
&&-\frac{\hat d_i}{\|s_i\|+\mu_i(t)}s_i\label{eq:wconELwoRV_ctr1}\\
\dot{\widehat{\Theta}}_i&=&-\Lambda_iY_i^T(q_i,\dot
q_i,\zerobf_p,\dot q_{ri})s_i\label{eq:wconELwoRV_ctr2}\\
\dot{\hat k}_i &=&\gamma_i s_i^Ts_i \label{eq:wconELwoRV_ctr3}\\
\dot{\hat d}_i &=&\delta_i\frac{\|s_i\|^2}{\|s_i\|+\mu_i(t)}\label{eq:wconELwoRV_ctr4}
\end{eqnarray}
\end{subequations}
where $\hat k_i$ is the time-varying control gain with $k_i(0)\geq 0$, $\Lambda_i$ is symmetric positive-definite, $\gamma_i$ is a positive constant, $\dot q_{ri}$ and $s_i$ are defined as in \eqref{eq:wconEL_auxivari2} and \eqref{eq:wconEL_auxivari3}, respectively,
$\widehat{\Theta}_i$, $\hat d_i$, and $\mu_i(t)$ are defined
as in \eqref{eq:wconEL_ctr}, and $Y_i(q_i,\dot q_i,\zerobf_p,{\dot q}_{ri})$ is defined as in (A3). Specifically, $Y_i(q_i,\dot q_i,\zerobf_p,{\dot q}_{ri})\Theta_i=C_i(q_i,\dot q_i)\dot q_{ri}+g_i(q_i)$.

Without loss of generality, we assume that the Laplacian matrix
$\Lcal_A$ has the form of \eqref{eq:coorEL-PFform1}.
We have the following main result without relative velocity measurements under a fixed directed graph.

\begin{theorem}\label{thm:wconELwoRV}
Suppose that the directed graph $\Gcal$ contains a directed spanning tree and the Laplacian matrix
$\Lcal_A$ has the form of \eqref{eq:coorEL-PFform1}. Using \eqref{eq:wconELwoRV_ctr} for
\eqref{eq:EL-system}, $\|q_i(t)-q_j(t)\|\to 0$ and $\dot
q_i(t)\to \zerobf_p$ as $t\to\infty$. In particular, $\lim_{t\to\infty} q_i(t)=\sum_{i=1}^{r_1} \frac{\xi_i}{\alpha_i} q_i(0)/\sum_{i=1}^{r_1} \frac{\xi_i}{\alpha_i}$, $i=1,\ldots, n$, where $\xi_i$ is defined as in Lemma \ref{lemma:spanningtree}, and $\alpha_i$ is defined in \eqref{eq:wconEL_auxivari1}.
\end{theorem}


\proof
By considering the form \eqref{eq:coorEL-PFform1} of the Laplacian matrix, we divide the agents into two sets. One is the set containing all roots and the other containing all non-root agents. The consensus convergence of all roots whose associated graph is strongly connected is studied first. And then the consensus convergence of the other agents is tackled through a leader-following framework. The final consensus equilibrium point depends only on the initial states of the roots. 

Using \eqref{eq:wconELwoRV_ctr1},
 the closed-loop system \eqref{eq:EL-system} can be written as
\begin{align}
    M_i(q_i)\dot{s}_i=&-C_i(q_i,\dot q_i)s_i-\hat k_i s_i-Y_i(q_i,\dot q_i,\zerobf_p,
{\dot q}_{ri})\widetilde{\Theta}_i\notag\\
&-M_i(q_i)\ddot q_{ri}-d_i(t)-\frac{\hat d_i}{\|s_i\|+\mu_i(t)}s_i\label{eq:wconELwoRV_closeloop1}
\end{align}
where $\widetilde{\Theta}_i\defeq \Theta_i-\widehat\Theta_i$.

We first consider the consensus convergence of the agents associated with $L_{11}$.
Consider the following positive weight function
\begin{align}\label{eq:wconELwoRV-Lya1}
V_{11}=&\frac{1}{2}\sum_{i=1}^{r_1} \Big[s_i^TM_i(q_i)s_i + \widetilde\Theta_i^T\Lambda_i^{-1}\widetilde\Theta_i+\gamma_i^{-1}(\hat k_i-\bar k_1)^2\notag\\
&+\delta_i^{-1}(\hat d_i-d_{\max})^2\Big]
\end{align}
where $\bar k_1$ is a positive constant to be determined later.
The derivative of $V_{11}(t)$ along \eqref{eq:wconELwoRV_closeloop1} can be written as
\begin{align}\label{eq:wconELwoRV-Lya1dot}
\dot V_{11}
=&\sum_{i=1}^{r_1}\Big[-\bar k_1 s_i^Ts_i-s_i^TM_i(q_i)\ddot q_{ri}-s_i^Td_i(t)\notag\\
&-\frac{\hat d_i\|s_i\|^2}{\|s_i\|+\mu_i(t)}+(\hat d_i-d_{\max})\frac{\|s_i\|^2}{\|s_i\|+\mu_i(t)}\Big]\notag\\
\leq&\sum_{i=1}^{r_1}\Big[-\bar k_1 s_i^Ts_i-s_i^TM_i(q_i)\ddot q_{ri}+d_{\max}\mu_i(t)\Big]
\end{align}

Define
$M(\bar q_1)\defeq \mbox{diag}(M_1(q_1),\cdots, M_{r_1}(q_{r_1}))$, $\tilde\xi_i\defeq \frac{\xi_i}{\alpha_i}/\sum_{i=1}^{r_1} \frac{\xi_i}{\alpha_i}$, $\widetilde\Xi_1\defeq \mbox{diag}(\tilde\xi_1,\cdots,\tilde\xi_{r_1})$, $\tilde q_i\defeq q_i-\sum_{j=1}^{r_1}\tilde\xi_jq_j$, $\Delta_1\defeq\mbox{diag}(\alpha_1,\cdots,\alpha_{r_1})$, and $\widetilde L_1\defeq \Delta_1L_{11}$. Let $\bar\vartheta_1$, $\bar q_1$, $\bar q_{r1}$, $\bar{\tilde q}_1$, $\bar s_1$, and $\bar{\tilde\xi}_1$ be the column stack vectors of, respectively, $\vartheta_i$, $q_i$, $q_{ri}$, $\tilde q_i$, $s_i$, and $\tilde\xi_i$, $i=1,\ldots, r_1$. We then get from \eqref{eq:wconEL_auxivari2} and \eqref{eq:wconEL_auxivari3} that
\begin{align}\label{eq:wconELwoRV_qdot}
\dot {\bar q}_1 =&~ \bar s_1-(\widetilde L_1\otimes I_p)\bar{\tilde q}_1-\int_0^t\bar\vartheta_1(\tau)\mbox{d}\tau
\end{align}
and
\begin{align*}
\ddot{\bar q}_{r1}=&~-(\widetilde L_1\otimes I_p)\dot{\bar q}_1-\bar\vartheta_1\notag\\
=&-[(\widetilde L_1+I_{r_1})\otimes I_p]\big(\bar s_1-\int_0^t\bar\vartheta_1(\tau)\mbox{d}\tau\big)+(\widetilde L_1\otimes I_p)^2\bar{\tilde q}_1.
\end{align*}
We can then get that
\begin{align}\label{eq:wconELwoRV_qrddotbound}
&\bar s_1^TM(\bar q_1)\ddot{\bar q}_{r1}\notag\\
\leq &~ k_{\overline m} \|\bar s_1\|\|\ddot{\bar q}_{r1}\| \notag\\
\leq &~k_{\overline m} \sigma_{\max}(\widetilde L_1+I_{n})\|\bar s_1\|^2+k_{\overline m}\sigma_{\max}^2(\widetilde L_1)\|\bar s_1\|\|\bar{\tilde q}_1\|\notag\\
&+k_{\overline m} \sigma_{\max}(\widetilde L_1+I_{n})\|s\|\|\int_0^t\bar\vartheta_1(\tau)\mbox{d}\tau\|
\end{align}
where we have used the fact
that $x^TPy\leq\sigma_{\max}(P)\|x\|\|y\|$, for vectors
$x$, $y$, and matrix $P$ with appropriate dimensions, to get the last inequality.

Consider the following positive weight function
\begin{align}\label{eq:wconELwoRV_Lya2}
V_{12}(t)=\sum_{i=1}^{r_1}\tilde\xi_i\tilde q_i^T \tilde q_i+\frac{\beta_1}{2}\Big[\int_0^t\bar\vartheta_1(\tau)\mbox{d}\tau\Big]^T\int_0^t\bar\vartheta_1(\tau)\mbox{d}\tau,
\end{align}
where $\beta_1$ is a positive constant to be determined later. Its derivative can be written as
\begin{align}\label{eq:wconELwoRV_Lya2dot}
\dot V_{12}(t)=&~\dot{\bar{\tilde q}}_1^T(\widetilde\Xi_1\otimes I_p)\bar{\tilde q}_1+{\bar{\tilde q}}_1^T(\widetilde\Xi_1\otimes I_p)\dot{\bar{\tilde q}}_1\notag\\
&+\beta_1\Big[\int_0^t\bar\vartheta_1(\tau)\mbox{d}\tau\Big]^T\bar\vartheta_1.
\end{align}
Note that
\begin{align}\label{eq:wconELwoRV_qtdot}
\dot{\bar{\tilde q}}_1=&\dot{\bar q}_1 -\onebf_{r_1}\otimes[(\tilde\xi^T\otimes I_p)\dot{\bar q}_1] \notag\\
=&-(\widetilde L_1\otimes I_p)\bar{\tilde q}_1+\bar s_1-\int_0^t\bar\vartheta_1(\tau)\mbox{d}\tau\notag\\
&- \onebf_{r_1}\otimes[(\tilde\xi^T\otimes I_p)(\bar s_1-\int_0^t\bar\vartheta_1(\tau)\mbox{d}\tau)]
\end{align}
where we have used \eqref{eq:wconELwoRV_qdot} and the fact that $\tilde\xi^T\widetilde L_1=\zerobf_{r_1}$ to obtain the second equality. Also note that
\begin{align}\label{eq:wconELwoRV_smv}
&(\widetilde\Xi_1\otimes I_p)\big\{\onebf_{r_1}\otimes[(\tilde\xi^T\otimes I_p)(\bar s_1-\int_0^t\bar\vartheta_1(\tau)\mbox{d}\tau)]\big\}\notag\\
=&\big\{[\widetilde\Xi_1(\onebf_{r_1}\otimes\tilde\xi^T)]\otimes I_p\big\}(\bar s_1-\int_0^t\bar\vartheta_1(\tau)\mbox{d}\tau)\notag\\
=& [(\tilde\xi\tilde\xi^T)\otimes I_p](\bar s_1-\int_0^t\bar\vartheta_1(\tau)\mbox{d}\tau).
\end{align}

Substituting \eqref{eq:wconELwoRV_qtdot} and \eqref{eq:wconELwoRV_smv} into \eqref{eq:wconELwoRV_Lya2dot}, we obtain
\begin{align}\label{eq:wconELwoRV_Lya2dot2}
\dot V_{12}(t)=&-\bar{\tilde q}_1^T[(\widetilde\Xi_1\widetilde L_1+\widetilde L_1^T\widetilde\Xi_1)\otimes I_p]\bar{\tilde q}_1-\beta_1\|\int_0^t\bar\vartheta_1(\tau)\mbox{d}\tau\|^2\notag\\
&+2{\bar{\tilde q}_1}^T[(\widetilde\Xi_1-\tilde\xi\tilde\xi^T)\otimes I_p](\bar s_1-\int_0^t\bar\vartheta_1(\tau)\mbox{d}\tau)\notag\\
&+\beta_1 \bar s_1^T\int_0^t\bar\vartheta_1(\tau)\mbox{d}\tau.
\end{align}

Define $\widetilde B_1\defeq \widetilde\Xi_1\widetilde L_1+\widetilde L_1^T\widetilde\Xi_1$. Note that $(\tilde\xi^T\otimes I_p)\tilde q=\zerobf_p$ and the subgraph associated with $L_{11}$ is strongly connected. It follows from Lemma \ref{lemma:dir-undir}
that
$\bar{\tilde q}_1^T(\widetilde B_1\otimes I_p)\bar{\tilde q}_1 \geq a(\widetilde B_1)\|\bar{\tilde q}_1\|^2$,
where $a(\widetilde B_1)>0$ is defined the same as in \eqref{eq:boundvec}. Since $\onebf_{r_1}^T\tilde\xi=1$, we can conclude that the matrix $\widetilde\Xi_1-\tilde\xi\tilde\xi^T$
is diagonally dominant and thus symmetric positive semidefinite. From Gersgorin Theorem, we can get that
$\sigma_{\max}(\widetilde\Xi_1-\tilde\xi\tilde\xi^T)\leq \frac{1}{2}.$
It follows from \eqref{eq:wconELwoRV_Lya2dot2} that
\begin{align}\label{eq:wconELwoRV_Lya2dot3}
\dot V_{12}(t)\leq &-a(\widetilde B_1)\|\bar{\tilde q}_1\|^2-\beta_1\|\int_0^t\bar\vartheta_1(\tau)\mbox{d}\tau\|^2+\|\bar{\tilde q}_1\|\|\bar s_1\|\notag\\
&+\|\bar{\tilde q}_1\|\|\int_0^t\bar\vartheta_1(\tau)\mbox{d}\tau\|+\beta_1 \|\bar s_1\|\|\int_0^t\bar\vartheta_1(\tau)\mbox{d}\tau\|.
\end{align}
Let
$\beta_1=\frac{1}{2a(\widetilde B_1)}+1$.
Note that
\begin{align*}
\|\bar{\tilde q}_1\|\|\int_0^t\bar\vartheta_1(\tau)\mbox{d}\tau\|\leq \frac{a(\widetilde B_1)}{2}\|\bar{\tilde q}_1\|^2 +\frac{1}{2a(\widetilde B_1)}\|\int_0^t\bar\vartheta_1(\tau)\mbox{d}\tau\|^2.
\end{align*}
Then we can obtain that
\begin{align}\label{eq:wconELwoRV_Lya2dot4}
\dot V_{12}(t)\leq &-\frac{a(\widetilde B_1)}{2}\|\bar{\tilde q}_1\|^2-\|\int_0^t\bar\vartheta_1(\tau)\mbox{d}\tau\|^2+\|\bar{\tilde q}_1\|\|\bar s_1\|\notag\\
&+\beta_1 \|\bar s_1\|\|\int_0^t\bar\vartheta_1(\tau)\mbox{d}\tau\|.
\end{align}

We then consider the following Lyapunov function candidate
\begin{align}\label{eq:wconELwoRV_Lya}
V_1(t)=V_{11}(t)+V_{12}(t).
\end{align}
From \eqref{eq:wconELwoRV-Lya1dot}, \eqref{eq:wconELwoRV_qrddotbound}, and \eqref{eq:wconELwoRV_Lya2dot4}, we obtain
\begin{align*}
\dot V_1(t)\leq &-\bar k_1\|\bar s_1\|^2 +k_{\overline m} \sigma_{\max}(\widetilde L_{11}+I_{r_1})\|\bar s_1\|^2\notag\\
&+\big[k_{\overline m} \sigma_{\max}(\widetilde L_{11}+I_{r_1})+\beta_1\big]\|\bar s_1\|\|\int_0^t\bar\vartheta_1(\tau)\mbox{d}\tau\|\notag\\
&+[k_{\overline m}\sigma_{\max}^2(\widetilde L_{11})+1]\|\bar s_1\|\|\bar{\tilde q}_1\|\notag\\
&-\frac{a(\widetilde B_1)}{2}\|\bar{\tilde q}_1\|^2-\|\int_0^t\bar\vartheta_1(\tau)\mbox{d}\tau\|^2+d_{\max}\sum_{i=1}^{r_1}\mu_i(t).
\end{align*}
Note that
\begin{align*}
&[k_{\overline m} \sigma_{\max}(\widetilde L_{11}+I_{r_1})+\beta_1]\|\bar s_1\|\|\int_0^t\bar\vartheta_1(\tau)\mbox{d}\tau\|\notag\\
\leq & \frac{[k_{\overline m} \sigma_{\max}(\widetilde L_{11}+I_{r_1})+\beta_1]^2}{2}\|\bar s_1\|^2+\frac{1}{2}\|\int_0^t\bar\vartheta_1(\tau)\mbox{d}\tau\|^2,
\end{align*}
and
\begin{align*}
  &[k_{\overline m}\sigma_{\max}^2(\widetilde L_{11})+1]\|\bar s_1\|\|\bar{\tilde q}_1\|\notag\\
  \leq & \frac{[k_{\overline m}\sigma_{\max}^2(\widetilde L_{11})+1]^2}{a(\widetilde B_1)}\|\bar s_1\|^2 + \frac{a(\widetilde B_1)}{4}\|\bar{\tilde q}_1\|^2.
\end{align*}
Choose $\bar k_1$ such that
\begin{align}\label{eq:wconELwoRV_kbar}
  \bar k_1=& k_{\overline m} \sigma_{\max}(\widetilde L_{11}+I_{r_1})+\frac{[k_{\overline m} \sigma_{\max}(\widetilde L_{11}+I_{r_1})+\beta_1]^2}{2}\notag\\
  &+\frac{[k_{\overline m}\sigma_{\max}^2(\widetilde L_{11})+1]^2}{a(\widetilde B_1)}+k_0,
\end{align}
with $k_0$ being a positive constant. We then have
\begin{align}\label{eq:wconELwoRV_Lyadot2}
\dot V_1(t)\leq &-k_0\|\bar s_1\|^2 -\frac{a(\widetilde B_1)}{4}\|\bar{\tilde q}_1\|^2-\frac{1}{2}\|\int_0^t\bar\vartheta_1(\tau)\mbox{d}\tau\|^2\notag\\
&+d_{\max}\sum_{i=1}^{r_1}\mu_i(t).
\end{align}

Integrating both sides of \eqref{eq:wconELwoRV_Lyadot2} and make some manipulation, we can get
\begin{align}\label{eq:wconELwoRV_Lya2dotf}
&V_1(t)+\frac{a(\widetilde B_1)}{4}\int_0^t\|\bar{\tilde q}_1(\tau)\|^2\mbox{d}\tau+\frac{1}{2}\int_0^t\|\int_0^{\tau}\bar\vartheta_1(x)\mbox{d}x\|^2\mbox{d}\tau\notag\\
&+k_0\int_0^t\|\bar s_1(\tau)\|^2\mbox{d}\tau
\leq V_1(0)+d_{\max}\sum_{i=1}^{r_1}\int_0^t\mu_i(\tau)\mbox{d}\tau.
\end{align}
Since $k_0, a(\widetilde B_1)>0$ and $\int_0^t\mu_i(\tau)\mbox{d}\tau<\infty$, $\forall i=1,\ldots,r_1$, we can get that $V(t)\in\mathbb{L}_{\infty}$, which means that $s_i, \widetilde\Theta_i, \hat k_i, \tilde q_i, \int_0^t\vartheta_i(\tau)\mbox{d}\tau\in \mathbb{L}_{\infty}$, and $s_i,\tilde q_i,\int_0^t\vartheta_i(\tau)\mbox{d}\tau\in\mathbb{L}_2$, $i=1,\ldots,r_1$.
From \eqref{eq:wconEL_auxivari1} and \eqref{eq:wconEL_auxivari3}, we can get $\vartheta_i, \dot q_i\in\mathbb{L}_2$, $i=1,\ldots,r_1$.

We then consider the agents associated with $L_{22}$.
Note that the eigenvalues of $\Lcal_A$ are the eigenvalues of $L_{11}$ and $L_{22}$. Under the condition that $\Gcal$ contains a directed spanning tree,
we can get from Lemma \ref{lemma:spanningtree} that $\Lcal_A$ has a single zero eigenvalue and all other eigenvalues
have positive real parts. Since the subgraph associated with $L_{11}$ is strongly connected, we can conclude that all the eigenvalues of $L_{22}$ have positive real parts. From Definition \ref{def:M-matrix}, we have $L_{22}$ is a nonsingular $M$-matrix.

Define
$M(\bar q_2)\defeq \mbox{diag}(M_{r_1+1}(q_{r_1+1}),\cdots, M_{n}(q_{n}))$, $\tilde q_i\defeq q_i-\sum_{j=1}^{r_1}\tilde\xi_jq_j$, $\Delta_2\defeq\mbox{diag}(\alpha_{r_1+1},\cdots,\alpha_n)$, and $\widetilde L_{22}\defeq \Delta_2L_{22}$. Let $\bar\vartheta_2$, $\bar q_2$, $\bar q_{r2}$, $\bar{\tilde q}_2$, and $\bar s_2$ be the column stack vectors of, respectively, $\vartheta_i$, $q_i$, $q_{ri}$, $\tilde q_i$, and $s_i$, $i=r_1+1,\ldots, n$. Since $L_{22}$ is a nonsingular $M$-matrix, $\widetilde L_{22}$ is also a nonsingular $M$-matrix. It follows from Lemma \ref{lemma:M-matrix} that there exists a diagonal matrix $\widetilde W_2\defeq \mbox{diag}(\tilde w_{r_1+1},\cdots, \tilde w_{n})$ with $\tilde w_i>0$, $i=r_1+1,\ldots,n$, such that $\widetilde Q\defeq \widetilde L_{22}^T\widetilde W_2+\widetilde W_2\widetilde L_{22}$ is positive definite. Define $\tilde w_{\max}\defeq \max_{i}\{w_{r_1+1},\ldots, w_n\}$. Motivated by the previous results, we consider the following Lyapunov function candidate
\begin{align}
V_2(t)=&\frac{1}{2}\sum_{i=r_1+1}^{n} \Big[s_i^TM_i(q_i)s_i + \widetilde\Theta_i^T\Lambda_i^{-1}\widetilde\Theta_i+\frac{(\hat k_i-\bar k_2)^2}{\gamma_i}\notag\\
&+w_i\tilde q_i^T \tilde q_i+\frac{(\hat d_i-d_{\max})^2}{\delta_i}\Big]\notag\\&+\frac{\beta_2}{2}\Big[\int_0^t\bar\vartheta_2(\tau)\mbox{d}\tau\Big]^T\int_0^t\bar\vartheta_2(\tau)\mbox{d}\tau
\end{align}
where $\bar k_2$ and $\beta_2$ are positive constants to be determined later. Its derivative can be written as
\begin{align}\label{eq:wconELwoRV-Lya2dot}
\dot V_{2}=&\sum_{i=r_1+1}^{n}\Big[-\bar k_2 s_i^Ts_i-s_i^TM_i(q_i)\ddot q_{ri}-\frac{\hat d_i\|s_i\|^2}{\|s_i\|+\mu_i(t)}\notag\\
&-s_i^Td_i(t)+(\hat d_i-d_{\max})\frac{\|s_i\|^2}{\|s_i\|+\mu_i(t)}\Big]+\dot{\bar{\tilde q}}_2^T(\widetilde W_2\otimes I_p){\bar{\tilde q}}_2\notag\\
&+{\bar{\tilde q}}_2^T(\widetilde W_2\otimes I_p)\dot{\bar{\tilde q}}_2+\beta_2\Big[\int_0^t\bar\vartheta_2(\tau)\mbox{d}\tau\Big]^T\bar\vartheta_2\notag\\
\leq&\sum_{i=r_1+1}^{n}\Big[-\bar k_2 s_i\!^Ts_i-s_i^TM_i(q_i)\ddot q_{ri}+d_{\max}\mu_i(t)\Big]\notag\\
&+\dot{\bar{\tilde q}}_2^T(\widetilde W_2\otimes I_p){\bar{\tilde q}}_2+{\bar{\tilde q}}_2^T(\widetilde W_2\otimes I_p)\dot{\bar{\tilde q}}_2\notag\\&+\beta_2\Big[\int_0^t\bar\vartheta_2(\tau)\mbox{d}\tau\Big]^T\vartheta_2.
\end{align}

Note that from \eqref{eq:wconEL_auxivari1} and \eqref{eq:wconEL_auxivari3},
\begin{align}\label{eq:wconELwoRV-Lya1dotiq0}
\bar s_2=\dot{\bar q}_2+(\widetilde L_{21}\otimes I_p)\bar{\tilde q}_1+(\widetilde L_{22}\otimes I_p)\bar{\tilde q}_2+\int_0^t\bar\vartheta_2(\tau)\mbox{d}\tau
\end{align}
where $\widetilde L_{21}\defeq \Delta_2L_{21}$.
Therefore, we have
\begin{align}\label{eq:wconELwoRV-Lya1dotiq1}
&\dot{\bar{\tilde q}}_2^T(\widetilde W_2\otimes I_p){\bar{\tilde q}}_2+{\bar{\tilde q}}_2^T(\widetilde W_2\otimes I_p)\dot{\bar{\tilde q}}_2\notag\\
=&2[\dot{\bar q}_2-\onebf_{r_2}\otimes\sum_{j=1}^{r_1}\tilde \xi_j\dot q_j]^T(\widetilde W_2\otimes I_p){\bar{\tilde q}}_2\notag\\
=&-{\bar{\tilde q}}_2^T(\widetilde Q\otimes I_p){\bar{\tilde q}}_2+2{\bar{\tilde q}}_2^T(\widetilde W_2\otimes I_p)\Big[\bar s_2-(\widetilde L_{21}\otimes I_p)\bar{\tilde q}_1\notag\\
&-\int_0^t\bar\vartheta_2(\tau)\mbox{d}\tau-\onebf_{r_2}\otimes\sum_{j=1}^{r_1}\tilde \xi_j\dot q_j\Big]\notag\\
\leq&-\lambda_{\min}(\widetilde Q)\|{\bar{\tilde q}}_2\|^2 +2\tilde w_{\max}\|{\bar{\tilde q}}_2\|\Big{\|}\bar s_2-(\widetilde L_{21}\otimes I_p)\bar{\tilde q}_1\notag\\
&-\int_0^t\bar\vartheta_2(\tau)\mbox{d}\tau-\onebf_{r_2}\otimes\sum_{j=1}^{r_1}\tilde \xi_j\dot q_j\Big{\|}\notag\\
\leq&-\frac{\lambda_{\min}(\widetilde Q)}{2}\|{\bar{\tilde q}}_2\|^2+\frac{8\tilde w_{\max}^2}{\lambda_{\min}(\widetilde Q)}\Big{(}\|\bar s_2\|^2+\sigma_{\max}^2(\widetilde L_{21})\|\bar{\tilde q}_1\|^2\notag\\
&+\|\int_0^t\bar\vartheta_2(\tau)\mbox{d}\tau\|^2+\|\onebf_{r_2}\otimes\sum_{j=1}^{r_1}\tilde \xi_j\dot q_j\|^2\Big{)}
\end{align}
where we have used the Cauchy-Schwarz Inequality to obtain the last inequality.
From \eqref{eq:wconEL_auxivari2} and \eqref{eq:wconEL_auxivari3},
\begin{align}\label{eq:wconELwoRV-Lya1dotiq2}
&\sum_{i=r_1+1}^ns_i^TM_i(q_i)\ddot q_{ri}\notag\\
=&-\bar s_2^TM(\bar q_2)\big[(\widetilde L_{21}\otimes I_p)\dot{\bar q}_1+(\widetilde L_{22}\otimes I_p)\dot{\bar q}_2-\bar\vartheta_2\big]\notag\\
\leq&~k_{\overline m}\sigma_{\max}(\widetilde L_{21})\|\bar s_2\|\|\dot{\bar q}_1\|+k_{\overline m}\|\bar s_2\|\|\bar\vartheta_2\|+k_{\overline m}\|\bar s_2\|\notag\\
&\Big{\|}(\widetilde L_{22}\otimes I_p)\big[{\bar s}_2-(\widetilde L_{21}\otimes I_p)\bar{\tilde q}_1-(\widetilde L_{22}\otimes I_p)\bar{\tilde q}_2\notag\\&-\int_0^t\bar\vartheta_2(\tau)\mbox{d}\tau)\big]\Big{\|}\notag\\
\leq&\frac{k_{\overline m}\sigma_{\max}(\widetilde L_{21})}{2}(\|\bar s_2\|^2+\|\dot{\bar q}_1\|^2)+k_{\overline m}\sigma_{\max}(\widetilde L_{22})\|\bar s_2\|^2\notag\\&+\frac{k_{\overline m}}{2}(3\|\bar s_2\|^2+2\big{\|}\int_0^t\bar\vartheta_2(\tau)\mbox{d}\tau\big{\|}^2)\notag\\
&+\frac{k_{\overline m}\sigma_{\max}(\widetilde L_{22}\widetilde L_{21})}{2}(\|\bar s_2\|^2+\|\bar{\tilde q}_1\|^2)\notag\\
&+\frac{k_{\overline m}^2\sigma_{\max}^4(\widetilde L_{22})}{\lambda_{\min}(\widetilde Q)}\|\bar s_2\|^2+\frac{\lambda_{\min}(\widetilde Q)}{4}\|\bar{\tilde q}_2\|^2\notag\\
&+\frac{k_{\overline m}\sigma_{\max}(\widetilde L_{22})}{2}(\|\bar s_2\|^2+\big{\|}\int_0^t\bar\vartheta_2(\tau)\mbox{d}\tau\big{\|}^2)
\end{align}
Also
from \eqref{eq:wconEL_auxivari3}, we have
\begin{align}\label{eq:wconELwoRV-Lya1dotiq3}
&\beta_2\Big[\int_0^t\bar\vartheta_2(\tau)\mbox{d}\tau\Big]^T\vartheta_2\notag\\
=&\beta_2\Big[\int_0^t\bar\vartheta_2(\tau)\mbox{d}\tau\Big]^T(\bar s_2-\int_0^t\bar\vartheta_2(\tau)\mbox{d}\tau)\notag\\
\leq&\frac{\beta_2}{2}\big{\|}\int_0^t\bar\vartheta_2(\tau)\mbox{d}\tau\big{\|}^2+\frac{\beta_2}{2}\|\bar s_2\|^2-\beta_2\big{\|}\int_0^t\bar\vartheta_2(\tau)\mbox{d}\tau\big{\|}^2\notag\\
=&-\frac{\beta_2}{2}\big{\|}\int_0^t\bar\vartheta_2(\tau)\mbox{d}\tau\big{\|}^2+\frac{\beta_2}{2}\|\bar s_2\|^2.
\end{align}

Choose $\beta_2$ and $\bar k_2$ to eliminate the terms associated with $\big{\|}\int_0^t\bar\vartheta_2(\tau)\mbox{d}\tau\big{\|}^2$ and $\|\bar s_2\|^2$ in (\ref{eq:wconELwoRV-Lya1dotiq1}-\ref{eq:wconELwoRV-Lya1dotiq3}), which yields
\begin{align}\label{eq:wconELwoRV-beta}
\beta_2=\frac{16\tilde w_{\max}^2}{\lambda_{\min}(\widetilde Q)}+2k_{\overline m}+k_{\overline m}\sigma_{\max}(\widetilde L_{22})+1
\end{align}
and
\begin{align}\label{eq:wconELwoRV-k}
\bar k_2=&k_{\overline m}\frac{\sigma_{\max}(\widetilde L_{21})+3+3\sigma_{\max}(\widetilde L_{22})+\sigma_{\max}(\widetilde L_{22}\widetilde L_{21})}{2}\notag\\
&+\frac{k_{\overline m}^2\sigma_{\max}^2(\widetilde L_{22})}{\lambda_{\min}(\widetilde Q)}+\frac{8\tilde w_{\max}^2}{\lambda_{\min}(\widetilde Q)}+\frac{\beta_2}{2}+k_0
\end{align}
with $k_0$ being a positive constant. Substituting (\ref{eq:wconELwoRV-Lya1dotiq1}-\ref{eq:wconELwoRV-k}) into \eqref{eq:wconELwoRV-Lya2dot}, we obtain
\begin{align}\label{eq:wconELwoRV-Lya2dotf}
\dot V_2\leq &-k_0\|\bar s_2\|^2-\frac{\lambda_{\min}(\widetilde Q)}{4}\|\bar{\tilde q}_2\|^2-\frac{1}{2}\big{\|}\int_0^t\bar\vartheta_2(\tau)\mbox{d}\tau\big{\|}^2\notag\\
&+\Phi(t)+d_{\max}\sum_{i=r_1+1}^n\mu_i(t)
\end{align}
where
\begin{align*}
\Phi(t)\defeq &\left(\frac{8\tilde w_{\max}^2\sigma_{\max}^2(\widetilde L_{21})}{\lambda_{\min}(\widetilde Q)}+\frac{k_{\overline m}\sigma_{\max}(\widetilde L_{22}\widetilde L_{21})}{2}\right)\|\bar{\tilde q}_1\|^2\notag\\
&\frac{8\tilde w_{\max}^2}{\lambda_{\min}(\widetilde Q)}\|\onebf_{r_2}\otimes\sum_{j=1}^{r_1}\tilde \xi_j\dot q_j\|^2+\frac{k_{\overline m}\sigma_{\max}(\widetilde L_{21})}{2}\|\dot{\bar q}_1\|^2.
\end{align*}
Since $\tilde q_i,\dot q_i\in\mathbb{L}_2$, $i=1,\ldots, r_1$, we have $\int_0^t\Phi(\tau)\mbox{d}\tau\in\mathbb{L}_{\infty}$. Integrating both sides of \eqref{eq:wconELwoRV-Lya2dotf} and after some manipulation, we can obtain
\begin{align}\label{eq:wconELwoRV-Lya2f}
&V_2(t)+\frac{\lambda_{\min}(\widetilde Q)}{4}\int_0^t\|\bar{\tilde q}_2(\tau)\|^2\mbox{d}\tau+\frac{1}{2}\int_0^t\big{\|}\int_0^{\tau}\bar\vartheta_2(x)\mbox{d}x\big{\|}^2\mbox{d}\tau\notag\\
&+\!k_0\int_0^t\!\|\bar s_2(\tau)\|^2\mbox{d}\tau\!\leq\! V_2(0)\!+\!\int_0^t\Phi(\tau)\mbox{d}\tau\!+\!d_{\max}\int_0^t\mu_i(\tau)\mbox{d}\tau.
\end{align}
Since $k_0,\lambda_{\min}(\widetilde Q)>0$ and $\int_0^t\mu_i(\tau)\mbox{d}\tau<\infty$, $i=r_1+1,\ldots,n$, we can get $V_2(t)\in\mathbb{L}_{\infty}$ and $\bar s_2$, $\bar{\tilde q}_2$, $\int_0^t\bar\vartheta_2(\tau)\mbox{d}\tau\in\mathbb{L}_2$.

Combining \eqref{eq:wconELwoRV_Lya2dotf} and \eqref{eq:wconELwoRV-Lya2f}, we obtain $s_i$, $\widetilde\Theta_i$, $\hat k_i$, $\tilde q_i$, $\int_0^t\vartheta_i(\tau)\mbox{d}\tau\in \mathbb{L}_{\infty}$, $i=1,\ldots, n$, and
$s_i$, $\tilde q_i$, $\int_0^t\vartheta_i(\tau)\mbox{d}\tau\in\mathbb{L}_2$, $i=1,\ldots,n$.
From \eqref{eq:wconEL_auxivari3}, we can get $\vartheta_i\in\mathbb{L}_{\infty}$ since $s_i,\int_0^t\vartheta_i(\tau)\mbox{d}\tau\in\mathbb{L}_{\infty}$, $i=1,\ldots,n$.
We can get from \eqref{eq:wconEL_auxivari1} and \eqref{eq:wconEL_auxivari2} that $\dot q_i, \ddot q_{ri}\in \mathbb{L}_{\infty}$, $i=1,\ldots,n$. Obviously, $\dot{\tilde q}_i\in\mathbb{L}_{\infty}$.
We then can get from (A1) and \eqref{eq:wconELwoRV_closeloop1} that $\dot s_i\in\mathbb{L}_{\infty}$, $i=1,\ldots,n$.

By far we get $\int_0^t\vartheta_i(\tau)\mbox{d}\tau, s_i, \tilde q_i\in\mathbb{L}_2\bigcap \mathbb{L}_{\infty}$ and $\vartheta_i, \dot s_i, \dot{\tilde q}_i\in\mathbb{L}_{\infty}$, it follows from Barbalat's Lemma that
$\lim_{t\to\infty} \|\int_0^t\vartheta_i(\tau)\mbox{d}\tau\|=0$, $\lim_{t\to\infty} \|s_i(t)\|=0$, and $\lim_{t\to\infty} \|\tilde q_i(t)\|=0$, $\forall i=1,\ldots, n$. It follows from \eqref{eq:wconEL_auxivari3} that $\lim_{t\to\infty} \|\vartheta_i(t)\|=0$, $\forall i=1,\ldots, n$. From the definition of $\tilde q_i$ and \eqref{eq:wconEL_auxivari1}, we can conclude that $\lim_{t\to\infty}\|q_i(t)-q_j(t)\|=0$ and $\lim_{t\to\infty}\|\dot q_i(t)\|=0$, $\forall i,j=1,\ldots,n$.

Following the same steps in the proof of Theorem \ref{thm:wconEL}, we can derive the consensus equilibrium, which is
\begin{align}
\lim_{t\to\infty} q_i(t)=\sum_{i=1}^{r_1} \frac{\xi_i}{\alpha_i} q_i(0)/\sum_{i=1}^{r_1} \frac{\xi_i}{\alpha_i}.
\end{align}
\endproof

\begin{remark}
Comparing with \eqref{eq:wconEL_ctr}, the control algorithm \eqref{eq:wconELwoRV_ctr} does not use the relative velocity information. However, the constant control gain $K_i$
in \eqref{eq:wconEL_ctr} is replaced with a time-varying adaptive gain $\hat k_i(t)$ in \eqref{eq:wconELwoRV_ctr}. It is intuitively true that the lack of information
would have additional requirements on the control gains. Although the choosing of $\bar k_1$, $\bar k_2$, $\beta_1$, and $\beta_2$ use some global information, they are only used for
the consensus convergence. The proposed algorithms \eqref{eq:wconEL_ctr} and \eqref{eq:wconELwoRV_ctr} are fully distributed in the sense that only the information of the agent and its neighbors are used, and there are no common gains among the agents.
\end{remark}

\subsection{Switching Directed Graphs}\label{sec:WconEL-wov-vary}

Although the derivative of $\dot q_{ri}$ is not used in the control algorithm, the consensus convergence analysis relies on the information of the Laplacian matrix, which cannot be used for switching directed graphs. Motivated by the model reference adaptive consensus scheme and the recent work in \cite{LiuJiRen16_Tcyber}, we propose the following reference model for each agent
\begin{align}\label{eq:wconEL-wo-sw-z}
  \ddot z_i & =-\sum_{j=1}^{n} a_{ij}(t)(q_i-q_j)-\Big(\frac{\sum_{j=1}^{n} a_{ij}(t)}{k_i}+k_i\Big)\dot q_i
\end{align}
where $k_i$ is a positive constant.
Define $e_i\defeq q_i-z_i$ and $w_i\defeq \dot e_i+e_i$. Note that $z_i$, $e_i$, and $w_i$ are redefined here.
Then the closed-loop of \eqref{eq:EL-system} can be written as
\begin{align}\label{eq:woconEL-sys-sw}
  M_i(q_i)\dot w_i\!+\!C_i(q_i,\dot q_i)w_i = & \tau_i\!-\!d_i\!-\!Y_i(q_i,\dot q_i,\ddot z_i\!-\!\dot e_i,\dot z_i\!-\!e_i).
\end{align}

To make $w_i\to 0$, we propose the following control algorithm
\begin{subequations}\label{eq:woconEL_ctr-sw}
\begin{eqnarray}
\tau_i &=&-K_{i}w_i+Y_i(q_i,\dot q_i,\ddot z_i-\dot e_i,\dot z_i-e_i)\widehat
\Theta_i\notag\\
&&-\frac{\hat d_i}{\|w_i\|+\mu_i(t)}w_i\label{eq:woconEL_ctr1-sw}\\
\dot{\widehat{\Theta}}_i&=&-\Lambda_iY_i(q_i,\dot q_i,\ddot z_i-\dot e_i,\dot z_i-e_i)w_i\label{eq:woconEL_ctr2-sw}\\
\dot{\hat d}_i&=&\delta_i\frac{\|w_i\|^2}{\|w_i\|+\mu_i(t)}\label{eq:woconEL_ctr3-sw}
\end{eqnarray}
\end{subequations}
where $K_i, \Lambda_i, \delta_i$, and $\mu_i(t)$ are defined the same as in \eqref{eq:wconEL_ctr}. We have the following result without relative velocity measurements under switching directed graphs.
\begin{theorem}\label{thm:woconEL-sw}
Suppose that the infinite sequence of switching graphs $\Gcal(t_i)$, $t=0,1,\ldots$ is uniformly jointly connected. Using \eqref{eq:woconEL_ctr-sw} for
\eqref{eq:EL-system}, $\|q_i(t)-q_j(t)\|\to 0$ and $\|\dot q_i(t)\|\to 0$ as $t\to\infty$.
\end{theorem}
\proof  Using \eqref{eq:woconEL_ctr-sw} for \eqref{eq:EL-system}, we have the following closed-loop system
\begin{align}\label{eq:woconEL_closeloop1-sw}
    M_i(q_i)\dot{w}_i=&-C_i(q_i,\dot q_i)w_i-Y_i(q_i,\dot q_i,\ddot z_i\!\!-\!\!\dot e_i,\dot z_i\!\!-\!\!e_i)\widetilde{\Theta}_i \notag\\
&-K_i w_i-d_i(t)-\frac{\hat d_i}{\|w_i\|+\mu_i(t)}w_i.
\end{align}

By considering the following Lyapunov function candidate
\begin{align*}
V=&~\frac{1}{2}\sum_{i=1}^n w_i^TM_i(q_i)w_i+\frac{1}{2}\sum_{i=1}^n\widetilde\Theta_i^T\Lambda_i^{-1}\widetilde\Theta_i\notag\\
&+\sum_{i=1}^n\frac{1}{2\delta_i}(\hat d_i-d_{\max}-1)^2
\end{align*}
and following the same steps in the proof of Theorem \ref{thm:wconEL-sw}, we can obtain
that $w_i,\widetilde\Theta_i,\hat d_i\in\mathbb{L}_\infty$ and $w_i\in\mathbb{L}_2\bigcap \mathbb{L}_1$.
From the definition of $w_i$, we can get that $e_i,\dot e_i\in\mathbb{L}_l$, $l\in[1,\infty]$.

Let $z$ and $e$ be, respectively, the stack vectors of $z_i$ and $e_i$, $i=1,\ldots,n$.
Define $y_{z_i}=z_i+(1/k_i)\dot z_i$. Set ${\xi_z} = {\left[ {z_1^T,...,z_n^T,y_{z_1}^T,...,y_{z_n}^T} \right]^T}$, ${e_z } = {\left[ e_1^T,...,e_n^T,\dot e_1^T,\ldots, \dot e_n^T \right]^T}$, and $K\defeq \mbox{diag}(k_1,\ldots,k_n)$. From the definition of $e_i$, \eqref{eq:wconEL-wo-sw-z} can be written in a vector form as
	\begin{align}\label{eq:FD2-vec-xi}
	\dot \xi_z  = -(L_1(t)\otimes I_p)\xi_z -(L_2(t)\otimes I_p) e_z
	\end{align}	
where
	\begin{align*}
	L_1(t)  =&~\left(
                              \begin{array}{cc}
                                K & -K \\
                                -K^{-1}\Acal(t) & K^{-1}\Dcal(t) \\
                              \end{array}
                            \right) \\
L_2(t)=&~\left(
                                    \begin{array}{cc}
                                      \zerobf_{n\times n} & \zerobf_{n\times n} \\
                                      K^{-1}\Lcal_A(t)  & K^{-2}\Dcal(t)+I \\
                                    \end{array}
                                  \right)
	\end{align*}	

From the analysis in \cite{LiuJiRen16_Tcyber}, $L_1(t)$ can be regarded as the Laplacian matrix of a directed graph $\Gcal_1(t)$ with $2n$ nodes. And if the sequence of switching graphs $\Gcal(t_i)$, $i=0,1,\ldots$ is uniformly jointly connected, the sequence of switching graphs $\Gcal_1(t_i)$, $i=0,1,\ldots$ is also uniformly jointly connected.
Since $L_2(t)$ is bounded, following the same analysis in the proof of Theorem \ref{thm:wconEL-sw}, we can get $\dot \xi_z, \xi_z\in\mathbb{L}_\infty$ from $e_z\in\mathbb{L}_1$.
From the definition of $\xi_z$, we can get $z_i,\dot z_i\in\mathbb{L}_\infty$. Since $e_i,\dot e_i\in\mathbb{L}_\infty$, $q_i,\dot q_i\in\mathbb{L}_\infty$. From \eqref{eq:wconEL-wo-sw-z}, $\ddot z_i\in\mathbb{L}_\infty$.
We then can get from (A1) and \eqref{eq:woconEL_closeloop1-sw} that $\dot w_i\in\mathbb{L}_\infty$. Combing with $w_i\in\mathbb{L}_2\bigcap\mathbb{L}_\infty$, we can conclude from Barbalat's Lemma that $\lim_{t\to\infty}\|w_i\|=0$. Since $w_i=\dot e_i+e_i$, we can get $\lim_{t\to\infty}\|e_i\|=\lim_{t\to\infty}\|\dot e_i\|=0$, which implies that $\lim_{t\to\infty}\|e_z\|=0$.
For the system \eqref{eq:FD2-vec-xi}, following the same steps in the proof of Theorem \ref{thm:wconEL}, we can get $\lim_{t\to\infty}\|z_i-z_j\|=0$ and $\lim_{t\to\infty}\|y_{z_i}-y_{z_j}\|=0$, which implies that $\lim_{t\to\infty}\|\dot z_i\|=0$.
Then we can conclude that $\lim_{t\to\infty}\|\dot q_i\|=0$ and $\lim_{t\to\infty}\|q_i-q_j\|=0$, $\forall i,j=1,\ldots,n$.
\endproof

\begin{remark}
  Compared with \cite{MeiRenChenMa13_automatica}, there are several differences. First, an integral term is introduced in the auxiliary variable design, and as a result, the Lyapunov function
candidate is redesigned by adding the term $\frac{\beta_1}{2}\Big[\int_0^t\bar\vartheta_1(\tau)\mbox{d}\tau\Big]^T\int_0^t\bar\vartheta_1(\tau)\mbox{d}\tau$, which is more difficult than that in \cite{MeiRenChenMa13_automatica}. Second, no common control gains are required in the proposed algorithm in this paper. Third, the external disturbances are not considered in \cite{MeiRenChenMa13_automatica} which are restrained by a robust continuous term. The differences in comparison with \cite{Wang14_TAC1} are the case without relative velocity feedback and the presence of external disturbances. More importantly, the consensus under switching directed graphs with very wild assumptions by using the model reference adaptive consensus scheme is also studied, which is not reported in \cite{MeiRenChenMa13_automatica} and \cite{Wang14_TAC1}.
\end{remark}

 \begin{remark}
 Just recently, the consensus for multiple Largrangian systems under switching directed graphs has been systematically solved in \cite{Wang17_CAC} by introducing a new novel sliding variable. With a different sliding variable, a consensus algorithm without using relative velocity information is proposed in \cite{Abdessameud18_ACC}. In this current paper, by using the model reference adaptive consensus scheme, both cases with and without relative velocity information are studied. One common feature of
the results in \cite{Wang17_CAC,Abdessameud18_ACC} and our work on switching graphs is that a system combined by the consensus for first-order integrators and a
vanishing term is obtained (Eq.(13) in \cite{Wang17_CAC}, Eq.(8) in \cite{Abdessameud18_ACC},
and Eqs. (31) and (40) in the current paper). The difference lies in
the vanishing term, which is the derivative of the sliding
variable in \cite{Wang17_CAC}, the sliding variable in \cite{Abdessameud18_ACC}, and the relative tracking error in the current paper. As a result, the novel
integral-input-output property of linear time-varying systems is introduced in \cite{Wang17_CAC} for the consensus convergence analysis, while the results in \cite{Abdessameud18_ACC} and this current paper only need the standard setting of input-output properties
of dynamical systems. The latter cannot be directly used for \cite{Wang17_CAC} as claimed therein.
Another difference is that the boundedness of the agents' positions $q_i$ has not been addressed in \cite{Abdessameud18_ACC},
where the results therein rely on the boundedness of the inertia matrix and the potential force. 
In Section \ref{sec:WconEL-wov-vary} of the current paper, the boundedness of all signals is addressed strictly, where the conclusion of $e_i\in\mathbb{L}_1$ plays an important role, with the help of the proposed robust term.
\end{remark}

\section{Simulation Results}\label{section_simulation}
For numerical simulations, we consider the leaderless consensus problem for six two-link revolute joint arms
modeled by Euler-Lagrange equations whose dynamics can be found in \cite[pp.123]{KellySantibanezLoria05}.
The additive disturbances are assumed to be $d_i(t)=0.2\sin(0.02it)$, $i=1,\ldots,6$. In particular, the masses of
links 1 and 2 are chosen as $m_{1i}=1\mbox{kg}$ and $m_{2i}=0.8\mbox{kg}$.
The lengths are $l_{1i}=0.8\mbox{m}$ and $l_{2i}=0.6~\mbox{m}$ and the distance from the previous joint to the center of mass are
$l_{c1i}=0.4\mbox{m}$ and $l_{c2i}=0.3~\mbox{m}$. The moments of inertia are $J_{1i}=0.0533\mbox{kg m}^2$ and $J_{2i}=0.024\mbox{kg m}^2$.
Let the initial angles of the six agents be,
respectively, $[-1,1]^T$, $[0,1]^T$, $[0.1,-1]^T$, $[-0.5,-1]^T$,
$[0,-0.5]^T$, and $[0.1,-0.5]^T$, and the initial angle derivatives be, respectively, $[-0.25, 0.25]^T$, $[-0.25, 0.15]^T$, $[0.02, 0.12]^T$,
$[-0.25,-0.15]^T$, $[0,-0.25]^T$, and $[0.15,0]^T$.

\subsection{Fixed Directed Graph}

\begin{figure}[!htb]
\centering
  \[\xymatrix{ & A_1  & A_5  & \\
 A_2 \ar@{<-}[r]\ar@{<->}[-1,1]\ar@{->}[-1,2] &  A_3 \ar@{<->}[u]\ar@{->}[r] & A_4
\ar@{<-}[u]\ar@{<->}[r]  & A_6\ar@{<-}[-1,-1]}\]
  \vspace{-.4cm}
  \caption{The directed graph that
characterizes the interaction among the six agents, where $A_i$,
$i=1,\ldots,6$, denotes the $i$th
agent.}\label{fig:c2nl-commugraph2}
\end{figure}
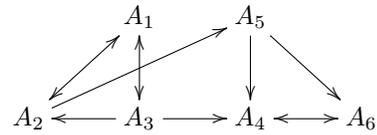

Fig. \ref{fig:c2nl-commugraph2} shows the fixed directed graph that
characterizes the interaction among the six agents. Note that the
directed graph contains a directed spanning tree and the subgraph associated with agents $A_1$, $A_2$, and $A_3$ is strongly connected.
The elements of the adjacency matrix are chosen as $a_{ij}=1$, if $A_j$ is a neighbor of $A_i$, and $a_{ij}=0$ otherwise.
Then we can compute the normalized left eigenvector of its associated Laplacian matrix with respect to the zero
eigenvalue is $\xi=[\frac{1}{3}, \frac{1}{6}, \frac{1}{2}, 0, 0,0]^T$.
From Theorem \ref{thm:wconEL} and \ref{thm:wconELwoRV}, the final states of the two links would be
$q(\infty)=\sum_{i=1}^n \frac{\xi_i}{\alpha_i} q_i(0)/\sum_{i=1}^n \frac{\xi_i}{\alpha_i}=[-0.2833, 0]^T$.

\begin{figure}[hhhhtb]
\centering
\begin{tabular}{cc}
          \subfigure[The angles]{
          \label{fig:wcEL-case1position}
          \includegraphics[width=.22\textwidth]{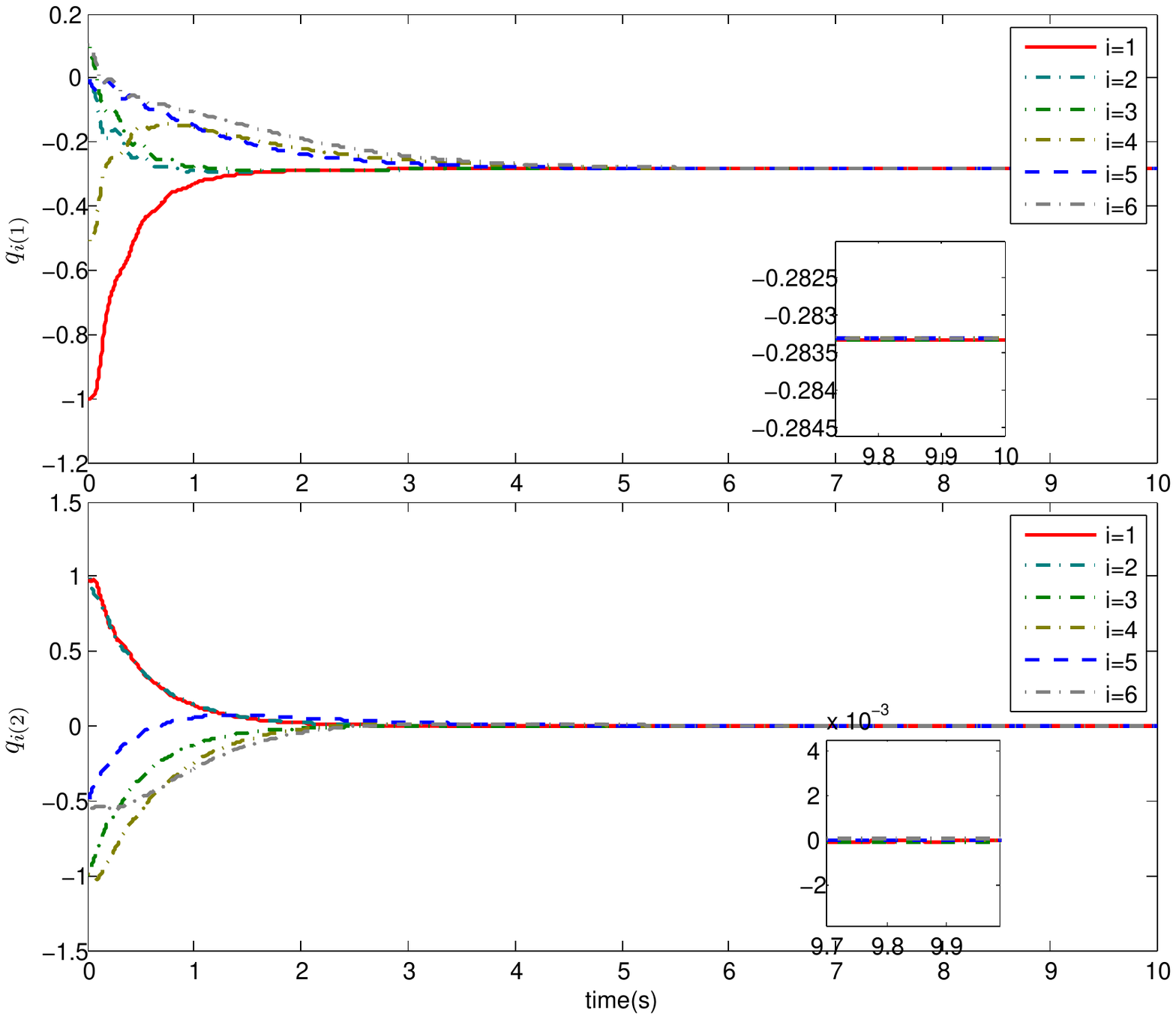}}
          &
      \subfigure[The angle derivatives]{
          \label{fig:wcEL-case1velocity}
          \includegraphics[width=.21\textwidth]{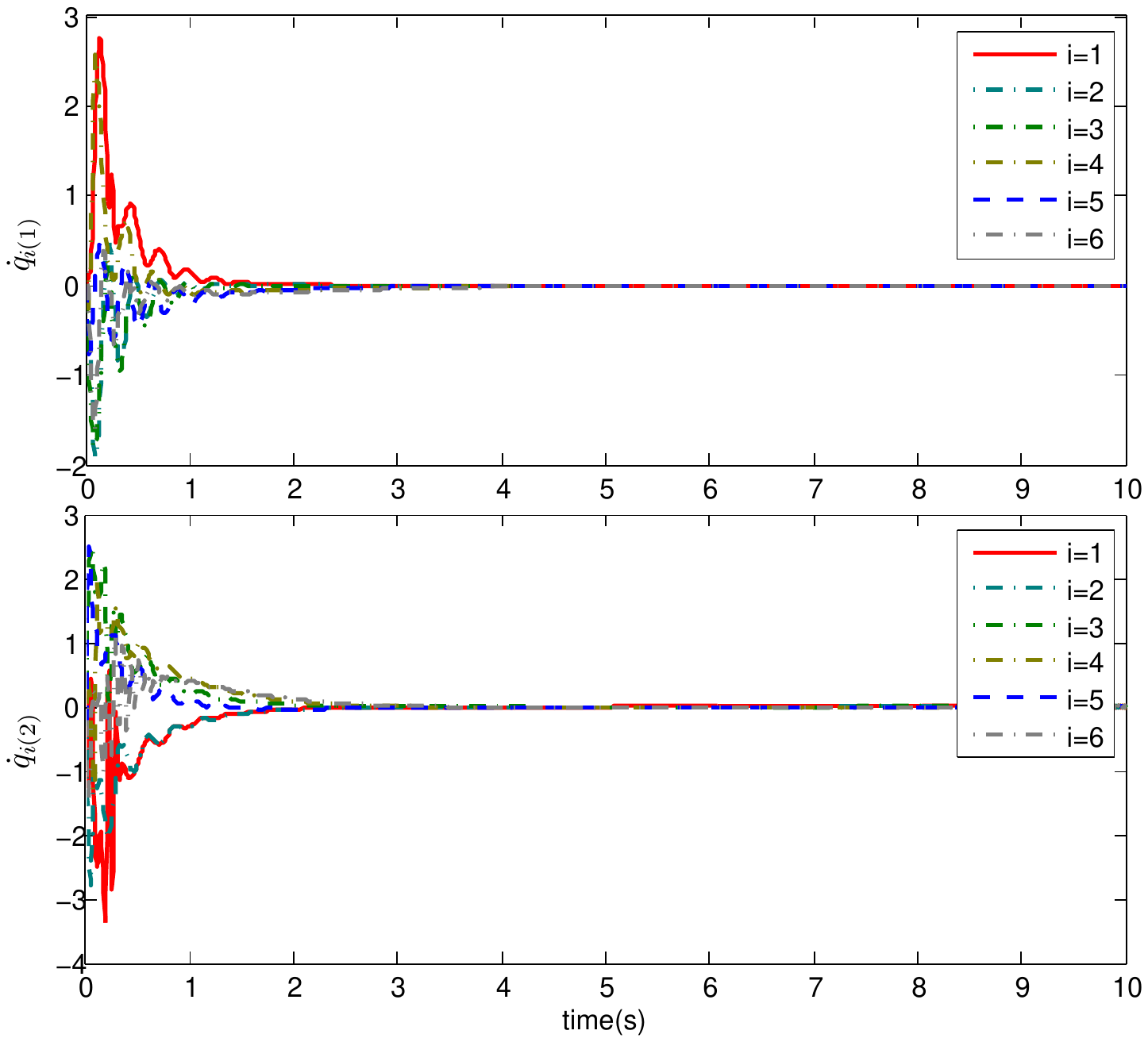}}
\end{tabular}
\caption{The angles and angle derivatives of the six agents using \eqref{eq:wconEL_ctr} under the
directed interaction graph in the presence of external disturbances.}
\end{figure}

%

For the control algorithm \eqref{eq:wconEL_ctr}, the control
parameters are chosen as $\alpha_i=1$, $K_i=2I_{2}$,
$\Lambda_i=5I_2$, and $\delta_i=0.2$, $i=1,\ldots,6$. The initial states of the
estimates $\widehat\Theta_i$ and $\hat d_i$ are all set to zero. $\mu_i$ is chosen as $e^{-t}$.
Fig.~\ref{fig:wcEL-case1position} and Fig.~\ref{fig:wcEL-case1velocity} show, respectively, the angles
and angle derivatives of the six agents using \eqref{eq:wconEL_ctr}. Clearly, the angles converge to the
weighted average of the initial angles of the agents, and the angle derivatives converge to zero.

%

\begin{figure}[hhhhtb]
\centering
\begin{tabular}{cc}
          \subfigure[The angles]{
          \label{fig:wcEL-case2position}
          \includegraphics[width=.22\textwidth]{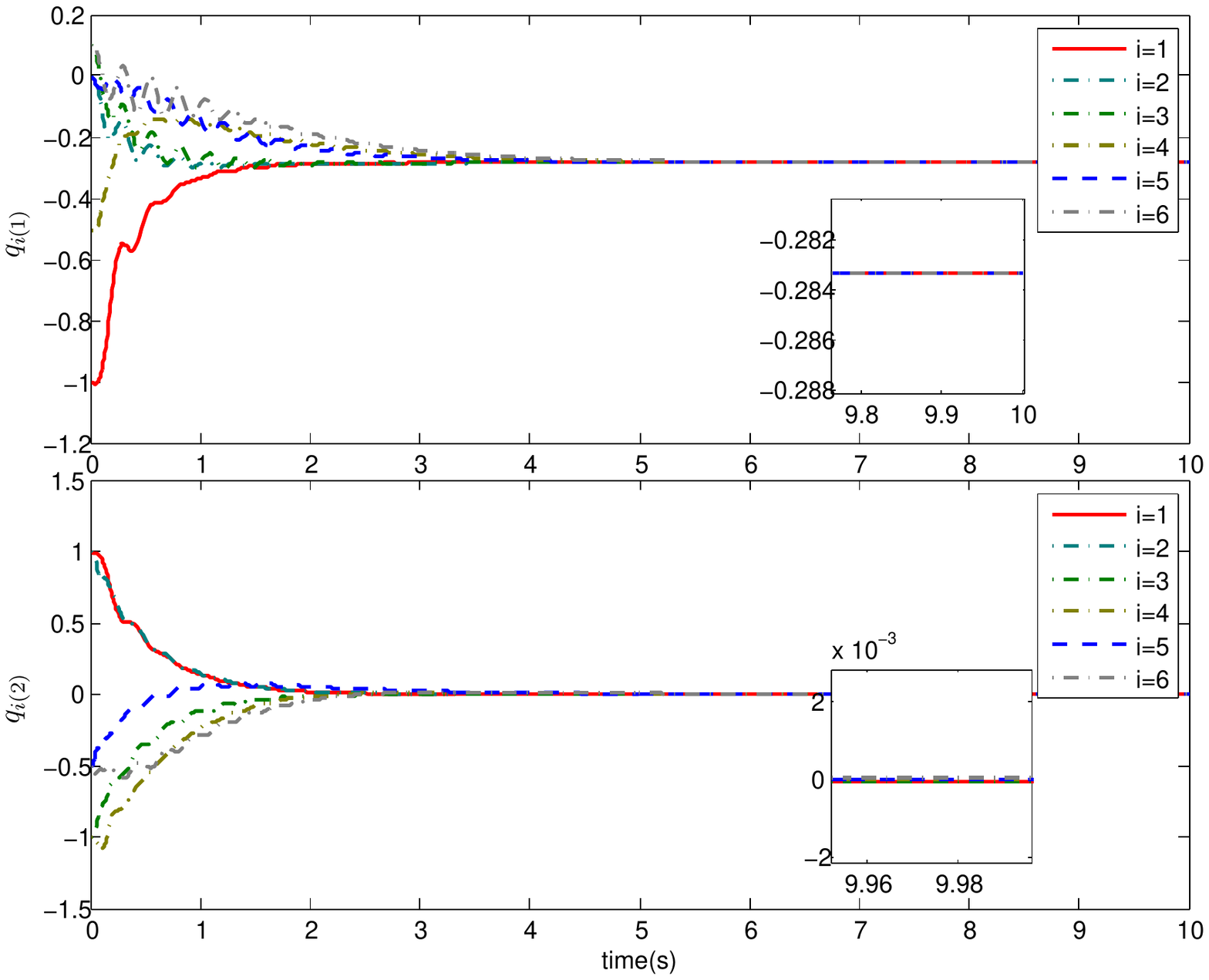}}
          &
      \subfigure[The angle derivatives]{
          \label{fig:wcEL-case2velocity}
          \includegraphics[width=.21\textwidth]{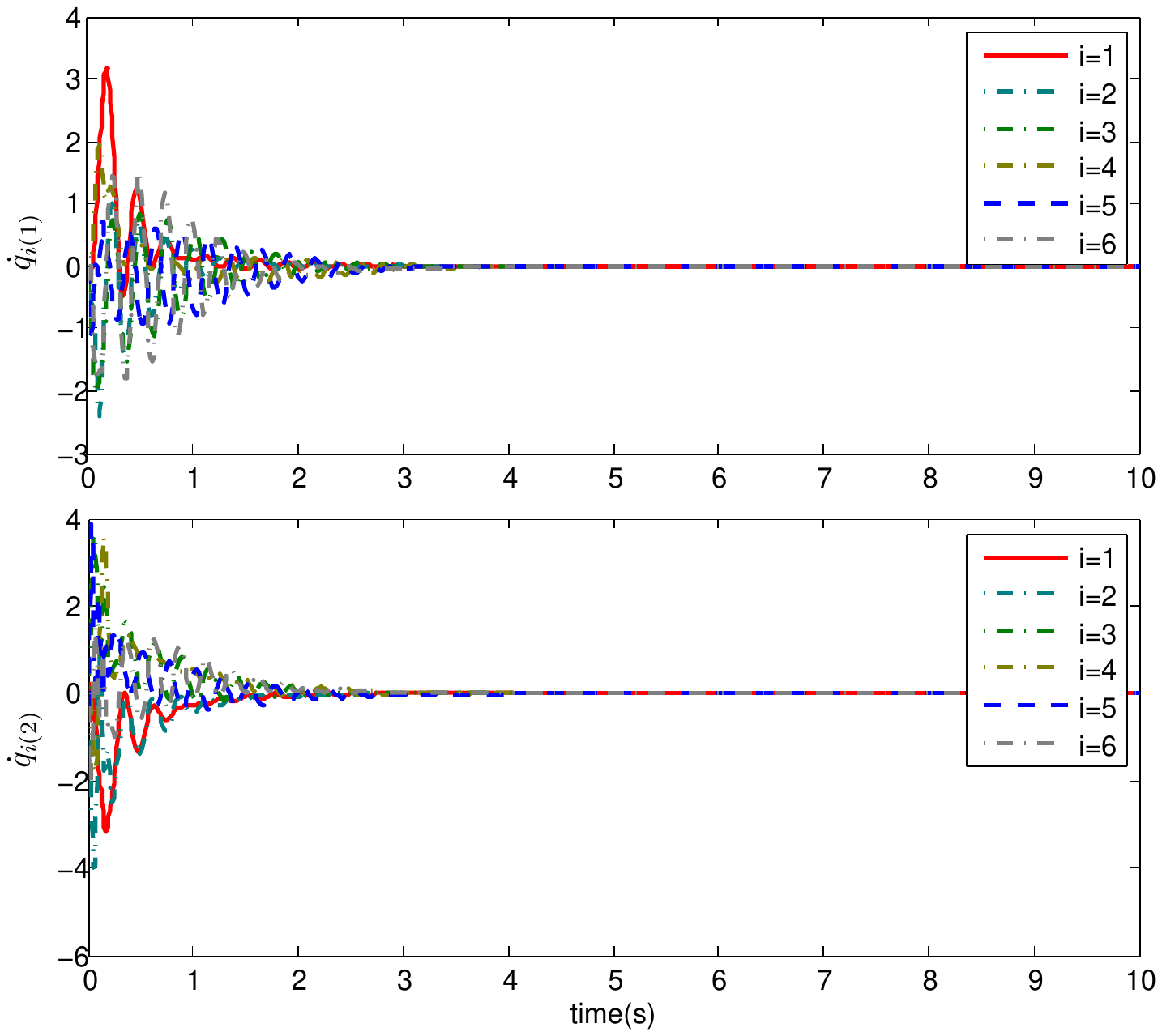}}
\end{tabular}
\caption{The angles and angle derivatives of the six agents using \eqref{eq:wconELwoRV_ctr} under the
directed interaction graph in the presence of external disturbances.}
\end{figure}

For the control algorithm \eqref{eq:wconELwoRV_ctr}, the control
parameters are chosen as $k_i=1$, $\Lambda_i=5I_2$, $\gamma_i=3$, and $\delta_i=0.2$, $i=1,\ldots,6$.
The initial states of the
estimates $\widehat\Theta_i$, $\hat d_i$, and $\hat k_i$ are all set to zero, and the initial states of $z_i(t)$ and $\dot z_i(t)$ are chosen as $z_i(0)=q_i(0)$ and $\dot z_i(t)=0$.
Fig.~\ref{fig:wcEL-case2position} and Fig.~\ref{fig:wcEL-case2velocity} show, respectively, the angles
and angle derivatives of the six agents using \eqref{eq:wconELwoRV_ctr}. It can be seen that the angles also converge to the
weighted average of the initial angles of the agents, and the angle derivatives converge to zero.

\subsection{Switching Directed Graphs}

\begin{figure}[!htb]
\centering
  \includegraphics[width=.4\textwidth]{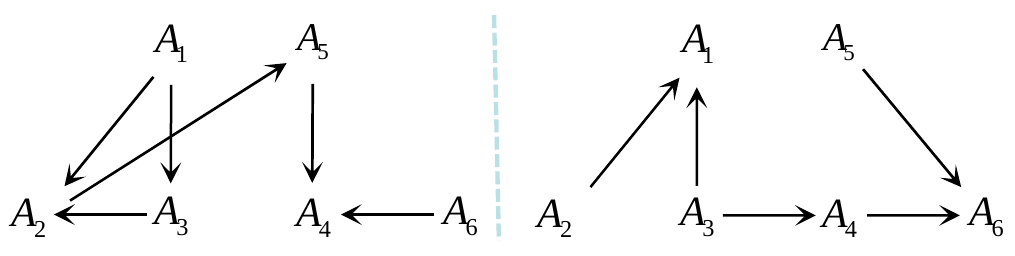}
  \vspace{-.4cm}
  \caption{The two possible directed graphs that
characterizes the interaction among the six agents.
}\label{fig:wcEL-graph-switch}
\end{figure}

Fig. \ref{fig:wcEL-graph-switch} shows the two possible directed graphs that characterizes the interaction among the six agents. None of the two graphs contains a directed spanning tree. However, the union of the two graphs is exactly the graph shown in Fig. \ref{fig:c2nl-commugraph2}. The entries of the Laplacian matrix and the initial states are chosen the same as before. And we allow that the underlying graphs switch between the two graphs in Fig. \ref{fig:wcEL-graph-switch} every two seconds.

%

\begin{figure}[hhhhtb]
\centering
\begin{tabular}{cc}
          \subfigure[The angles]{
          \label{fig:wcEL-case1position-sw}
          \includegraphics[width=.21\textwidth]{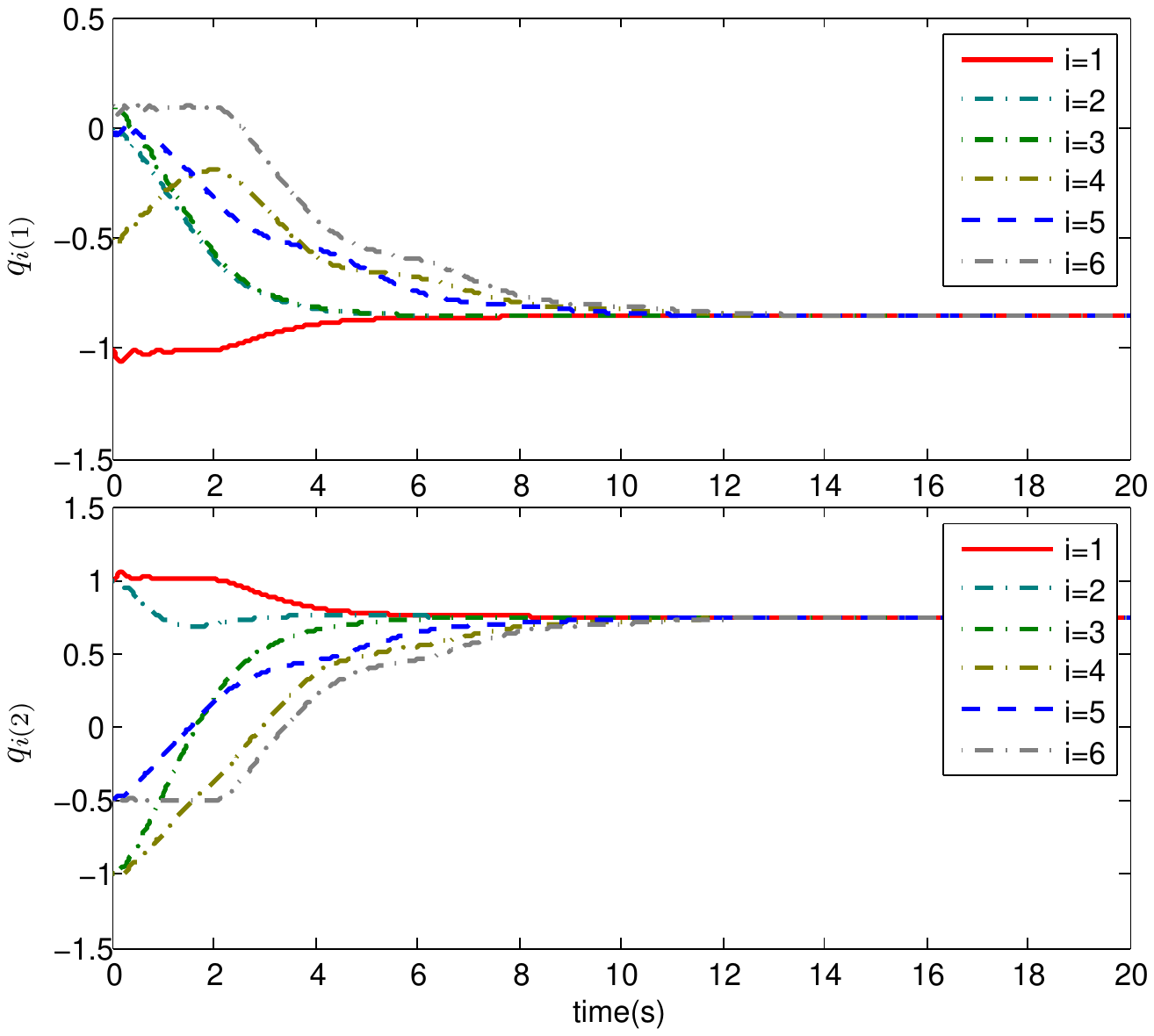}}
          &
      \subfigure[The angle derivatives]{
          \label{fig:wcEL-case1velocity-sw}
          \includegraphics[width=.21\textwidth]{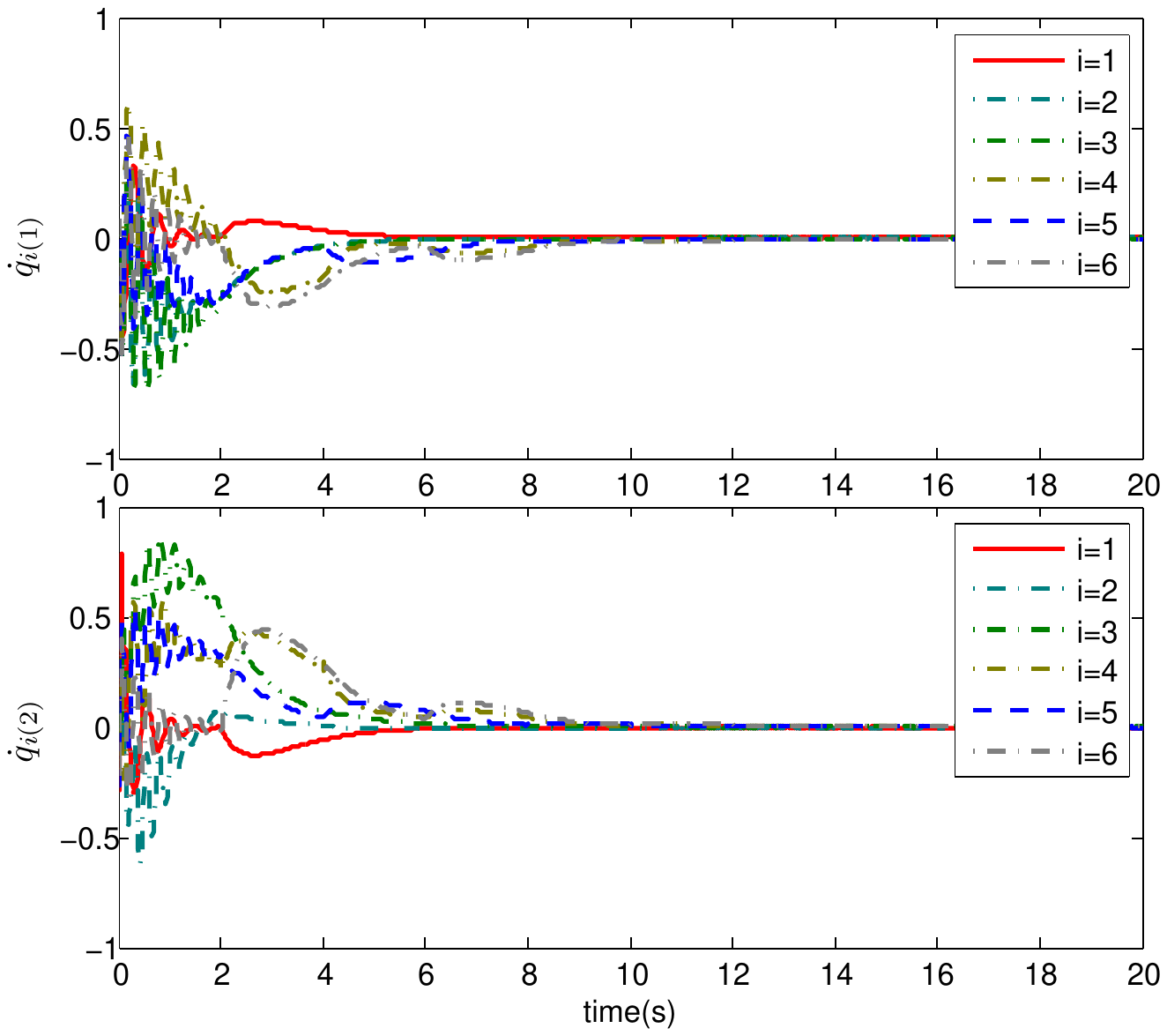}}
\end{tabular}
\caption{The angles and angle derivatives of the six agents using \eqref{eq:wconEL_ctr-sw} under switching directed interaction graphs in the presence of external disturbances.}
\end{figure}

For the control algorithm \eqref{eq:wconEL_ctr-sw}, the control
parameters are chosen as $\alpha_i=1$, $K_i=2I_{2}$,
$\Lambda_i=5I_2$, and $\delta_i=0.2$, $\forall i=1,\ldots,6$. The initial states of the
estimates $\widehat\Theta_i$ and $\hat d_i$ are all set to zero, and the initial states of $z_i(t)$ is chosen as $z_i(0)=q_i(0)$, $i=1,\ldots, 6$.
Fig.~\ref{fig:wcEL-case1position-sw} and Fig.~\ref{fig:wcEL-case1velocity-sw} show, respectively, the angles
and angle derivatives of the six agents using \eqref{eq:wconEL_ctr-sw}. Clearly, the angles converge to the same value and the angle derivatives converge to zero.

%

\begin{figure}[hhhhtb]
\centering
\begin{tabular}{cc}
          \subfigure[The angles]{
          \label{fig:wcEL-case2position-sw}
          \includegraphics[width=.23\textwidth]{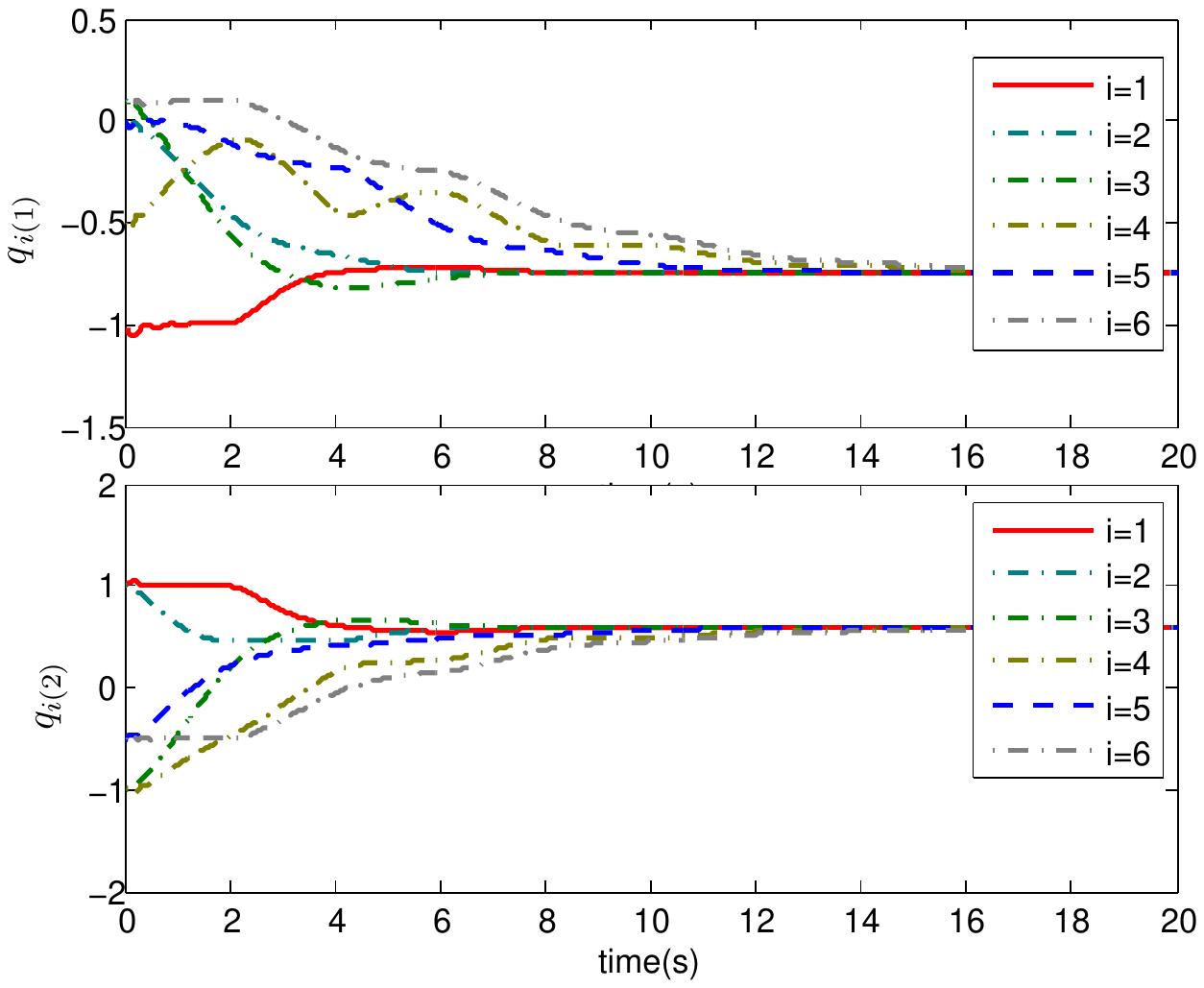}}
          &
      \subfigure[The angle derivatives]{
          \label{fig:wcEL-case2velocity-sw}
          \includegraphics[width=.23\textwidth]{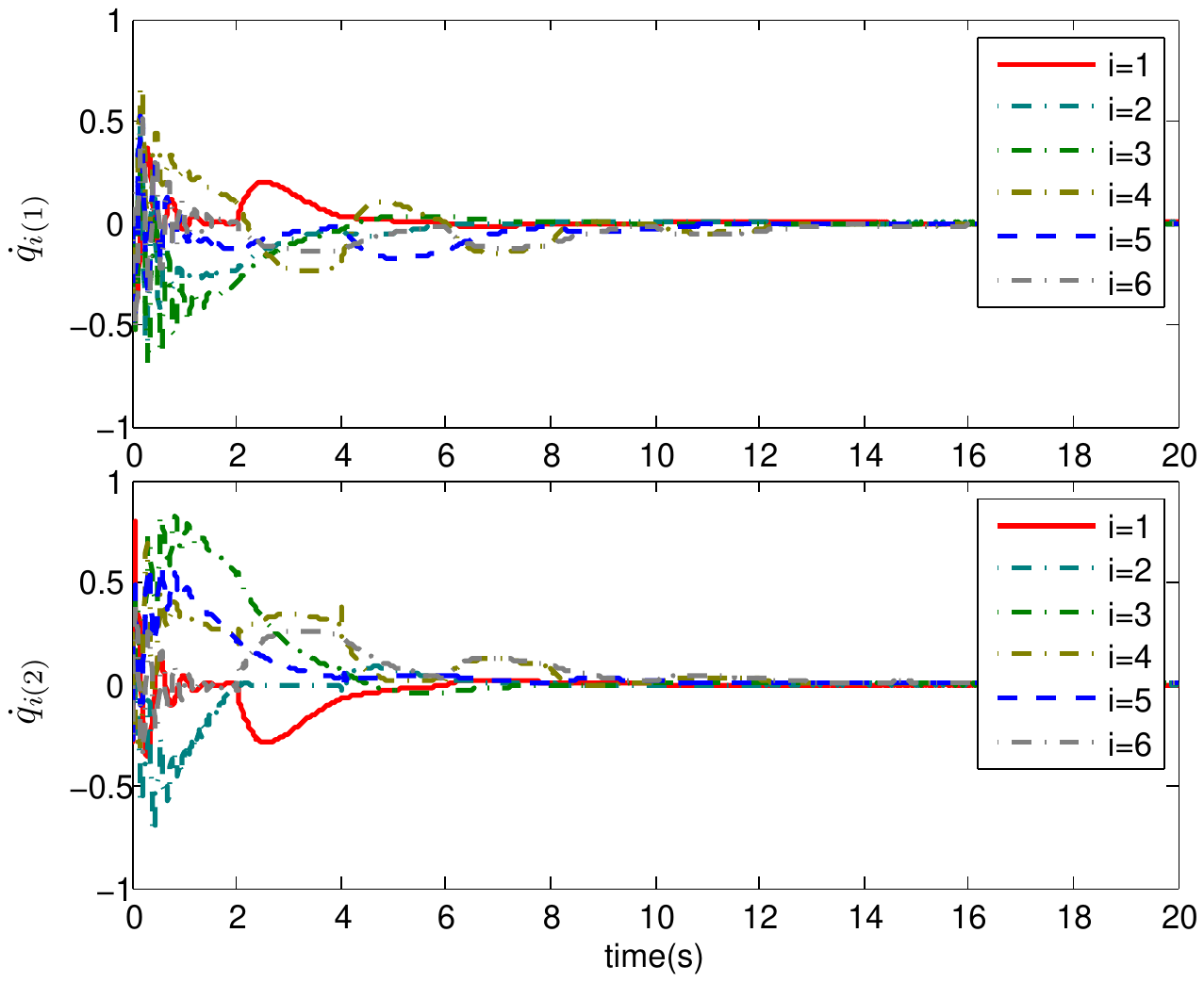}}
\end{tabular}
\caption{The angles and angle derivatives of the six agents using \eqref{eq:woconEL_ctr-sw} under switching directed interaction graphs in the presence of external disturbances.}
\end{figure}

For the control algorithm \eqref{eq:woconEL_ctr-sw}, the control
parameters are chosen as $\alpha_i=1$, $\Lambda_i=5I_2$, $\gamma_i=3$, and $\delta_i=0.2$, $\forall i=1,\ldots,6$.
The initial states of the
estimates $\widehat\Theta_i$, $\hat d_i$, and $\hat k_i$ are all set to zero.
Fig.~\ref{fig:wcEL-case2position-sw} and Fig.~\ref{fig:wcEL-case2velocity-sw} show, respectively, the angles
and angle derivatives of the six agents using \eqref{eq:woconEL_ctr-sw}. It can be seen that the angles also converge to the same constant and the angle derivatives converge to zero.

\section{Conclusions}\label{section_conclusion}

The leaderless consensus problem for
multiple Lagrangian systems in the presence of parametric uncertainties and external disturbances under a directed graph has been studied. We have considered both cases with and without using neighbors' velocity measurements. Asymptotic consensus convergence has been shown with the help of a robust continuous term with adaptive varying gains. For a fixed directed graph, with the introduction of an integral term in the auxiliary variable design,
the final consensus equilibrium of the systems has been explicitly derived. We have shown that this equilibrium is dependent on
three factors, namely, the interactive topology, the initial positions of the agents, and the control gains
of the proposed control algorithms. For switching directed graphs, a model reference adaptive consensus based method has been proposed for both the cases with and without relative velocity feedback.

\begin{IEEEbiography}[{\includegraphics[width=1in,height=1.25in,clip,keepaspectratio]{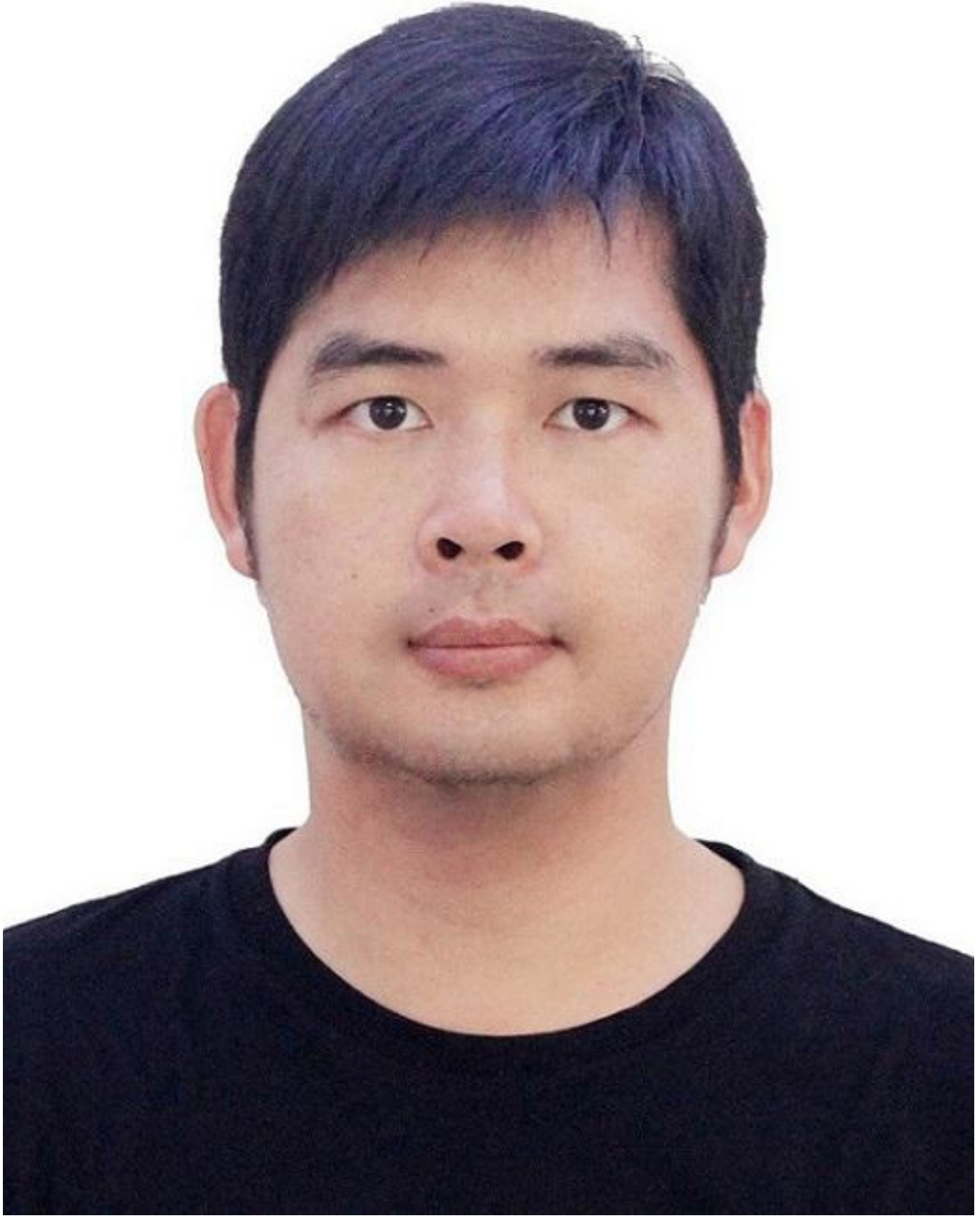}}]{Jie Mei} (M'14) received the B.S. degree in Information and Computing Science from Jilin University, Changchun, China, in 2007, and the Ph.D. degree in Control Science and Engineering from the Harbin Institute of Technology, Harbin, China, in 2011.

He was an exchange Ph.D. student supported by the China Scholarship Council with the Department of Electrical and Computer Engineering, Utah State University, Logan, UT, USA, from 2009 to 2011. He held postdoctoral research positions with the Harbin Institute of Technology Shenzhen Graduate School, Guangdong, China, the City University of Hong Kong, Hong Kong, and the University of California at Riverside, Riverside, CA, USA, from 2012 to 2015. Since June 2015, he has been with the School of Mechanical Engineering and Automation, Harbin Institute of Technology, Shenzhen, Guangdong, China, where he is currently an Associate Professor. His current research interests include coordination of distributed multi-agent systems and its applications on formation of unmanned vehicles and robots.
\end{IEEEbiography}


\begin{thebibliography}{10}
\providecommand{\url}[1]{#1}
\csname url@samestyle\endcsname
\providecommand{\newblock}{\relax}
\providecommand{\bibinfo}[2]{#2}
\providecommand{\BIBentrySTDinterwordspacing}{\spaceskip=0pt\relax}
\providecommand{\BIBentryALTinterwordstretchfactor}{4}
\providecommand{\BIBentryALTinterwordspacing}{\spaceskip=\fontdimen2\font plus
\BIBentryALTinterwordstretchfactor\fontdimen3\font minus
  \fontdimen4\font\relax}
\providecommand{\BIBforeignlanguage}[2]{{%
\expandafter\ifx\csname l@#1\endcsname\relax
\typeout{** WARNING: IEEEtran.bst: No hyphenation pattern has been}%
\typeout{** loaded for the language `#1'. Using the pattern for}%
\typeout{** the default language instead.}%
\else
\language=\csname l@#1\endcsname
\fi
#2}}
\providecommand{\BIBdecl}{\relax}
\BIBdecl

\bibitem{CaoYuRenChen13_TII}
Y.~Cao, W.~Yu, W.~Ren, and G.~Chen, ``An overview of recent progress in the
  study of distributed multi-agent coordination,'' \emph{{IEEE} Transactions on
  Industrial Informatics}, vol.~9, no.~1, pp. 427--438, February 2013.

\bibitem{KnornChenMiddleton16_TCNS}
S.~Knorn, Z.~Chen, and R.~H. Middleton, ``Overview: Collective control of
  multiagent systems,'' \emph{IEEE Transactions on Control of Network Systems},
  vol.~3, no.~4, pp. 334--347, Dec 2016.

\bibitem{QinMaShiWang17_TIE}
J.~Qin, Q.~Ma, Y.~Shi, and L.~Wang, ``Recent advances in consensus of
  multi-agent systems: A brief survey,'' \emph{IEEE Transactions on Industrial
  Electronics}, vol.~64, no.~6, pp. 4972--4983, June 2017.

\bibitem{OhParlAhn15_Automatica}
K.-K. Oh, M.-C. Park, and H.-S. Ahn, ``A survey of multi-agent formation
  control,'' \emph{Automatica}, vol.~53, no.~3, pp. 424--440, March 2015.

\bibitem{NedicOR18_PIEEE}
A.~Nedi$\acute{\mbox{c}}$, A.~Olshevsky, and M.~G. Rabbat, ``Network topology
  and communication-computation tradeoffs in decentralized optimization,''
  \emph{Proceedings of the IEEE}, vol. 106, no.~5, pp. 953--976, May 2018.

\bibitem{ScardoviArcakSontag10_TAC}
L.~Scardovi, M.~Arcak, and E.~Sontag, ``Synchronization of interconnected
  systems with applications to biochemical networks: An input-output
  approach,'' \emph{{IEEE} Transactions on Automatic Control}, vol.~55, no.~6,
  pp. 1367 --1379, June 2010.

\bibitem{ChenWen14_TAC}
W.~Chen, C.~Wen, S.~Hua, and C.~Sun, ``Distributed cooperative adaptive
  identification and control for a group of continuous-time systems with a
  cooperative pe condition via consensus,'' \emph{{IEEE} Transactions on
  Automatic Control}, vol.~59, no.~1, pp. 91--106, January 2014.

\bibitem{RenBeardAtkins07_CSM}
W.~Ren, R.~W. Beard, and E.~M. Atkins, ``Information consensus in multivehicle
  cooperative control,'' \emph{{IEEE} Control Systems Magazine}, vol.~27,
  no.~2, pp. 71--82, April 2007.

\bibitem{MeiRenChen16_TAC}
J.~Mei, W.~Ren, and J.~Chen, ``Distributed consensus of second-order
  multi-agent systems with heterogeneous unknown inertias and control gains
  under a directed graph,'' \emph{{IEEE} Transactions on Automatic Control},
  vol.~61, no.~8, pp. 2019--2034, August 2016.

\bibitem{LiuJiRen16_Tcyber}
K.~Liu, Z.~Ji, and W.~Ren, ``Necessary and sufficient conditions for consensus
  of second-order multiagent systems under directed topologies without global
  gain dependency,'' \emph{IEEE Transactions on Cybernetics}, vol.~47, no.~8,
  pp. 2089--2098, 2017.

\bibitem{ScardoviSepulchre10_automatica}
L.~Scardovi and R.~Sepulchre, ``Synchronization in networks of identical linear
  systems,'' \emph{Automatica}, vol.~45, no.~11, pp. 2557--2562, Novermeber
  2009.

\bibitem{LiDuanChenHuang10_TCS1}
Z.~Li, Z.~Duan, G.~Chen, and L.~Huang, ``Consensus of multi-agent systems and
  synchronization of complex networks: A unified viewpoint,'' \emph{{IEEE}
  Transactions on Circutis and Systems-1: Regular papers}, vol.~57, no.~1, pp.
  213--224, January 2010.

\bibitem{HouChengTan09_SMCB}
Z.~Hou, L.~Cheng, and M.~Tan, ``Decentralized robust adaptive control for the
  multiagent system consensus problem using neural networks,'' \emph{{IEEE}
  Transactions on Systems, Man, and Cybernetics-Part B: Cybernetics}, vol.~39,
  no.~3, pp. 636--647, 2009.

\bibitem{KellySantibanezLoria05}
R.~Kelly, V.~Santibanez, and A.~Loria, \emph{Control of Robot Manipulators in
  Joint Space}.\hskip 1em plus 0.5em minus 0.4em\relax London: Springer, 2005.

\bibitem{ChengHouTan07}
L.~Cheng, Z.~Hou, and M.~Tan, ``Decentralized adaptive consensus control for
  multi-manipulator system with uncertain dynamics,'' in \emph{Proceedings of
  IEEE International Conference on Systems, Man, and Cybernetics}, Singapore,
  2008, pp. 2712--2717.

\bibitem{Ren09_IJC}
W.~Ren, ``Distributed leaderless consensus algorithms for networked
  {E}uler-{L}agrange systems,'' \emph{International Journal of Control},
  vol.~82, no.~11, pp. 2137--2149, 2009.

\bibitem{NunoOBH11_TAC}
E.~Nuno, R.~Ortega, L.~Basanez, and D.~Hill, ``Synchronization of networks of
  nonidentical {E}uler-{L}agrange systems with uncertain parameters and
  communication delays,'' \emph{{IEEE} Transactions on Automatic Control},
  vol.~56, no.~4, pp. 935--941, April 2011.

\bibitem{MeiRenMa11_Automatica}
J.~Mei, W.~Ren, and G.~Ma, ``Distributed containment control for {L}agrangian
  networks with parametric uncertainties under a directed graph,''
  \emph{Automatica}, vol.~48, no.~4, pp. 653--659, April 2012.

\bibitem{MeiRenChenMa13_automatica}
J.~Mei, W.~Ren, J.~Chen, and G.~Ma, ``Distributed adaptive coordination for
  multiple {L}agrangian systems under a directed graph without using neighbors'
  velocity information,'' \emph{Automatica}, vol.~49, no.~6, pp. 1723--1731,
  2013.

\bibitem{ZhangTangHuang18_TII}
W.~Zhang, Y.~Tang, T.~Huang, and A.~V. Vasilakos, ``Consensus of networked
  {E}uler-{L}agrange systems under time-varying sampled-data control,''
  \emph{IEEE Transactions on Industrial Informatics}, vol.~14, no.~2, pp.
  535--544, Feb 2018.

\bibitem{LiuJia17_IJCAS}
Y.~Liu and Y.~Jia, ``Adaptive consensus control for multiple {E}uler-{L}agrange
  systems with external disturbance,'' \emph{International Journal of Control,
  Automation and Systems}, vol.~15, no.~1, pp. 205-- 211, 2017.

\bibitem{Wang14_TAC1}
H.~Wang, ``Consensus of networked mechanical systems with communication delays:
  A unified framework,'' \emph{{IEEE} Transactions on Automatic Control},
  vol.~59, no.~6, pp. 1571--1576, 2014.

\bibitem{Wang13_automatica}
------, ``Flocking of networked uncertain {Euler-Lagrange} systems on directed
  graphs,'' \emph{Automatica}, vol.~49, no.~9, pp. 2774--2779, 2013.

\bibitem{Wang17_CAC}
------, ``Dynamic feedback for consensus of networked {L}agrangian systems with
  switching topology,'' in \emph{2017 Chinese Automation Congress (CAC)}, Oct
  2017, pp. 1340--1345.

\bibitem{Abdessameud18_ACC}
A.~Abdessameud, ``Distributed consensus of euler-lagrange systems under
  switching directed graphs,'' in \emph{2018 Annual American Control Conference
  (ACC)}, June 2018, pp. 56--61.

\bibitem{YeAndersonYu17_IJRNC}
M.~Ye, B.~D. Anderson, and C.~Yu, ``Distributed model-independent consensus of
  {E}uler-{L}agrange agents on directed networks,'' \emph{International Journal
  of Robust and Nonlinear Control}, vol.~27, no.~14, pp. 2428--2450.

\bibitem{Mei15_CAC}
J.~Mei, ``Weighted consensus for multiple {L}agrangian systems under a directed
  graph,'' in \emph{2015 Chinese Automation Congress (CAC)}, 2015, pp.
  1064--1068.

\bibitem{Mei17_ACC}
------, ``Weighted consensus for multiple {L}agrangian systems under a directed
  graph without using neighbors’ velocity measurements,'' in
  \emph{Proceedings of the American Control Conference}, Seattle, USA, May
  24--May 26 2017, pp. 1353--1357.

\bibitem{HokayemStipanovicSpong09}
P.~F. Hokayem, D.~M. Stipanovic, and M.~W. Spong, ``Semiautonomous control of
  multiple networked {L}agrangian systems,'' \emph{International Journal of
  Robust and Nonlinear Control}, vol.~19, no.~18, pp. 2040--2055, 2009.

\bibitem{ChungSlotine09}
S.-J. Chung and J.-J.~E. Slotine, ``Cooperative robot control and concurrent
  synchronization of {L}agrangian systems,'' \emph{{IEEE} Transactions on
  Robotics}, vol.~25, no.~3, pp. 686--700, June 2009.

\bibitem{MengDimaJohan14_TRO}
Z.~Meng, D.~V. Dimarogonas, and K.~H. Johansson, ``Leader-follower coordinated
  tracking of multiple heterogeneous {L}agrange systems using continuous
  control,'' \emph{IEEE Transactions on Robotics}, vol.~30, no.~3, pp.
  739--745, June 2014.

\bibitem{CaiHuang16_TAC}
H.~Cai and J.~Huang, ``The leader-following consensus for multiple uncertain
  {E}uler-{L}agrange systems with an adaptive distributed observer,''
  \emph{IEEE Transactions on Automatic Control}, vol.~61, no.~10, pp.
  3152--3157, Oct 2016.

\bibitem{AbdessameudTayebi17_TAC}
A.~Abdessameud, A.~Tayebi, and I.~G. Polushin, ``Leader-follower
  synchronization of {E}uler-{L}agrange systems with time-varying leader
  trajectory and constrained discrete-time communication,'' \emph{IEEE
  Transactions on Automatic Control}, vol.~62, no.~5, pp. 2539--2545, May 2017.

\bibitem{YangFangChenJiangCao17_TAC}
Q.~Yang, H.~Fang, J.~Chen, Z.~P. Jiang, and M.~Cao, ``Distributed global
  output-feedback control for a class of {E}uler-{L}agrange systems,''
  \emph{IEEE Transactions on Automatic Control}, vol.~62, no.~9, pp.
  4855--4861, Sept 2017.

\bibitem{ChenDong17_IJRNC}
C.~Chen and W.~Dong, ``Distributed tracking control of uncertain mechanical
  systems with velocity constraints,'' \emph{International Journal of Robust
  and Nonlinear Control}, vol.~27, no.~17, pp. 3990--4012.

\bibitem{LiuYeQinYu18_TSMC}
Q.~Liu, M.~Ye, J.~Qin, and C.~Yu, ``Event-triggered algorithms for
  leader-follower consensus of networked {E}uler-{L}agrange agents,''
  \emph{IEEE Transactions on Systems, Man, and Cybernetics: Systems}, pp.
  1--13, 2017.

\bibitem{FengHuWenDixonMei18_TCNS}
Z.~Feng, G.~Hu, W.~Ren, W.~E. Dixon, and J.~Mei, ``Distributed coordination of
  multiple unknown {E}uler-{L}agrange systems,'' \emph{IEEE Transactions on
  Control of Network Systems}, vol.~5, no.~1, pp. 55--66, March 2018.

\bibitem{ChenSongGuan18_TNNLS}
G.~Chen, Y.~Song, and Y.~Guan, ``Terminal sliding mode-based consensus tracking
  control for networked uncertain mechanical systems on digraphs,'' \emph{IEEE
  Transactions on Neural Networks and Learning Systems}, vol.~29, no.~3, pp.
  749--756, March 2018.

\bibitem{KlotzObuzKanDixon18_TCyber}
J.~R. Klotz, S.~Obuz, Z.~Kan, and W.~E. Dixon, ``Synchronization of uncertain
  {E}uler-{L}agrange systems with uncertain time-varying communication
  delays,'' \emph{IEEE Transactions on Cybernetics}, vol.~48, no.~2, pp.
  807--817, Feb 2018.

\bibitem{ChopraStipanovicSpong08}
N.~Chopra, D.~M. Stipanovic, and M.~W. Spong, ``On synchronization and
  collision avoidance for mechanical systems,'' in \emph{Proceedings of the
  American Control Conference}, Seattle, Washington, June 2008, pp. 3713--3718.

\bibitem{GhapaniMeiRenSong16_Automatica}
S.~Ghapani, J.~Mei, W.~Ren, and Y.~Song, ``Fully distributed flocking with a
  moving leader for {L}agrange networks with parametric uncertainties,''
  \emph{Automatica}, vol.~67, pp. 67 -- 76, 2016.

\bibitem{MengRenYou10_automaitca}
Z.~Meng, W.~Ren, and Z.~You, ``Distributed finite-time attitude containment
  control for multiple rigid bodies,'' \emph{Automatica}, vol.~46, no.~12, pp.
  2092--2099, December 2010.

\bibitem{SlotineLi91}
J.-J.~E. Slotine and W.~Li, \emph{Applied Nonlinear Control}.\hskip 1em plus
  0.5em minus 0.4em\relax Englewood Cliffs, New Jersey: Prentice Hall, 1991.

\bibitem{RenBeard08}
W.~Ren and R.~W. Beard, \emph{Distributed Consensus in Multi-vehicle
  Cooperative Control}.\hskip 1em plus 0.5em minus 0.4em\relax London:
  Springer-Verlag, 2008.

\bibitem{BermanPlemmons79}
A.~Berman and R.~J. Plemmons, \emph{Nonnegative Matrices in the Mathematical
  Sciences}.\hskip 1em plus 0.5em minus 0.4em\relax New York: Academic Press,
  INC., 1979.

\bibitem{YucelenHaddad13_TAC}
T.~Yucelen and W.~Haddad, ``Low-frequency learning and fast adaptation in model
  reference adaptive control,'' \emph{{IEEE} Transactions on Automatic
  Control}, vol.~58, no.~4, pp. 1080--1085, 2013.

\bibitem{Mei18_CDC}
J.~Mei, ``Model reference adaptive consensus for uncertain multi-agent systems
  under directed graphs,'' in \emph{Proceedings of the {IEEE} Conference on
  Decision and Control}, FL, USA, 2018, pp. 6198--6203.

\bibitem{Rugh96}
W.~J. Rugh, \emph{Linear System Theory}, 2nd~ed.\hskip 1em plus 0.5em minus
  0.4em\relax Englewood Cliffs, New Jersey: Prentice Hall, 1996.

\end{thebibliography}
\end{document}